\documentclass[prd,showkeys,floatfix,twocolumn,amsmath,amssymb,floatfix]{revtex4}
\usepackage{graphicx}
\usepackage{epstopdf}
\usepackage{multirow}
\usepackage{subfigure}
\usepackage{color}
\usepackage[colorlinks,citecolor=blue,anchorcolor=red,menucolor=red,linkcolor=red,filecolor=red,runcolor=red,urlcolor=blue,frenchlinks=red]{hyperref}
\usepackage{enumitem}

\makeatletter
\renewcommand{\@thesubfigure}{\hskip\subfiglabelskip}
\makeatother
\allowdisplaybreaks[3]

\begin{document}
\title{Two- and three-gluon glueballs of $C=+$}
%

\author{Hua-Xing Chen$^1$}
\email{hxchen@seu.edu.cn}
\author{Wei Chen$^2$}
\email{chenwei29@mail.sysu.edu.cn}
\author{Shi-Lin Zhu$^3$}
\email{zhusl@pku.edu.cn}

\affiliation{
$^1$School of Physics, Southeast University, Nanjing 210094, China\\
$^2$School of Physics, Sun Yat-Sen University, Guangzhou 510275, China\\
$^3$School of Physics and Center of High Energy Physics, Peking University, Beijing 100871, China}

\begin{abstract}
We study two- and three-gluon glueballs of $C=+$ using the method of QCD sum rules. We systematically construct their interpolating currents, and find that all the spin-1 currents of $C=+$ vanish. This suggests that the ``ground-state'' spin-1 glueballs of $C=+$ do not exist within the relativistic framework. We calculate masses of the two-gluon glueballs with $J^{PC} = 0^{\pm+}/2^{\pm+}$ and the three-gluon glueballs with $J^{PC} = 0^{\pm+}/2^{\pm+}$. We propose to search for the $J^{PC} = 0^{-+}/2^{-\pm}/3^{\pm-}$ three-gluon glueballs in their three-meson decay channels in future BESIII, GlueX, LHC, and PANDA experiments.
\end{abstract}
\maketitle
\pagenumbering{arabic}

%
%
%
\section{Introduction}
\label{sec:intro}
%

Glueballs, composed of valence gluons, are important for the understanding of non-perturbative QCD~\cite{Freund:1975pn,Fritzsch:1975tx,Jaffe:1975fd}. There have been tremendous theoretical studies on them in the past fifty years using various models and methods, such as the MIT bag model~\cite{Chodos:1974je}, the flux-tube model~\cite{Isgur:1984bm}, the Coulomb Gauge model~\cite{Szczepaniak:1995cw,LlanesEstrada:2005jf}, Regge trajectories~\cite{Szanyi:2019kkn}, holographic QCD~\cite{Zhang:2021itx}, Lattice QCD~\cite{Wilson:1974sk,Chen:2005mg,Mathieu:2008me,Meyer:2004gx,Gregory:2012hu}, and QCD sum rules~\cite{Novikov:1979va,Novikov:1981xi,Kataev:1981aw,Kataev:1981gr,Narison:1984hu,Narison:1996fm,Bagan:1990sy,Latorre:1987wt,Forkel:2003mk,Hao:2005hu,Qiao:2014vva,Qiao:2015iea,Pimikov:2016pag,Pimikov:2017xap,Pimikov:2017bkk,Narison:2021xhc}, etc. However, experimental efforts in searching for glueballs are confronted with the difficulty of identifying them unambiguously, and there is currently no definite experimental evidence for their existence.

Recently the D0 and TOTEM Collaborations studied $pp$ and $p\bar p$~\cite{Abazov:2012qb} cross sections, which are found to be different with a significance of $3.4\sigma$~\cite{Abazov:2020rus}. Together with their previous result~\cite{Antchev:2017yns}, this significance can be increased to $5.2\sigma$--$5.7\sigma$. The above difference leads to the evidence of a $t$-channel exchanged odderon~\cite{Levin:1990gg,Braun:1998fs,Cudell:2002xe,Khoze:2017swe,Martynov:2018sga}, that is predominantly a three-gluon glueball of $C=-$. We refer to Refs.~\cite{Lukaszuk:1973nt,Bartels:1980pe,Kwiecinski:1980wb,Donnachie:1983ff,Khoze:2018bus,Csorgo:2018uyp,Goncalves:2018nsp,Xie:2019soz,Dumitru:2021tqp} and the review~\cite{Block:2006hy} for more discussions. Due to these studies, interests in glueballs are reviving recently. Since the above odderon evidence is still indirect, it is crucial and important to directly study the glueball itself.

The lowest-lying two-, three-, and four-gluon glueballs have been systematically investigated in Ref.~\cite{Jaffe:1985qp}, where the authors constructed their corresponding non-relativistic low-dimension operators. These operators have been successfully used in Lattice QCD calculations. In this paper we systematically study two- and three-gluon glueballs of $C=+$. We shall construct their corresponding relativistic glueball currents, and calculate masses of these glueballs using the method of QCD sum rules. The same approach has been applied in Ref.~\cite{Chen:2021cjr} to study three-gluon glueballs of $C=-$, so a rather complete QCD sum rule study will be done on the lowest-lying glueballs composed of two or three valence gluons. These studies can largely improve our understanding of the gluon degree of freedom as well as the non-perturbative behaviors of the strong interaction at the low energy region.

This paper is organized as follows. We systematically construct relativistic two- and three-gluon glueball currents of $C=+$ in Sec.~\ref{sec:current}. We apply them to perform QCD sum rule analyses in Sec.~\ref{sec:sumrule}, and perform numerical analyses in Sec.~\ref{sec:numerical}. The obtained results are summarized and discussed in Sec.~\ref{sec:summary}, which are compared with Lattice QCD results~\cite{Chen:2005mg,Mathieu:2008me,Meyer:2004gx,Gregory:2012hu}.

%
\section{Relativistic glueball currents}
\label{sec:current}
%

In this section we systematically construct relativistic glueball currents, including the two-gluon glueball currents and the $C=+$ three-gluon glueball currents. We shall do this separately in the following subsections. Note that the two-gluon glueball currents can not reach $C=-$~\cite{Yang:1950rg}, and the $C=-$ three-gluon glueball currents have been systematically constructed in Ref.~\cite{Chen:2021cjr}.

\subsection{Couplings of tensor currents}

In the present study we shall use some special tensor currents to study glueballs with non-zero spins $J \neq 0$. These currents have $2 \times J$ Lorentz indices with certain symmetries, and they couple to both positive- and negative-parity glueballs. In this subsection we briefly explain how we deal with them.

We assume $J_{\alpha\beta}$ to be a tensor current with two antisymmetric Lorentz indices $\mu$ and $\nu$. Taking the current $J_{\alpha\beta} = \bar c \sigma_{\alpha\beta} c$ as an example, it can be separated into ($\alpha,\beta=0,1,2,3$ and $i,j=1,2,3$):
\begin{equation}
J_{\alpha\beta} = \bar c \sigma_{\alpha\beta} c \rightarrow
\left\{\begin{array}{l}
\bar c \sigma_{ij} c \, , \, {P=+} \, ,
\\[1mm]
\bar c \sigma_{0i} c \, , \, {P=-} \, .
\end{array}\right.
\end{equation}
Accordingly, it couples to both positive- and negative-parity charmonia through
\begin{eqnarray}
\langle 0 | J_{\alpha\beta} | h_c (\epsilon,p) \rangle &=& i f^T_{h_c} \epsilon_{\alpha\beta\mu\nu} \epsilon^\mu p^\nu \, ,
\\ \langle 0 | J_{\alpha\beta} | J/\psi (\epsilon,p) \rangle &=& i f^T_{J/\psi} (p_\alpha\epsilon_\beta - p_\beta\epsilon_\alpha) \, ,
\end{eqnarray}
where $f^T_{h_c}$ and $f^T_{J/\psi}$ are relevant decay constants. Given the Lorentz structures of $J/\psi$ and $h_c$ to be totally different, they can be clearly separated from each other. For example, we can isolate $h_c$ at the hadron level by investigating the two-point correlation function containing
\begin{eqnarray}
&& \langle 0 | J_{\alpha\beta} | h_c \rangle \langle h_c | J_{\alpha^\prime\beta^\prime}^\dagger | 0 \rangle
\label{eq:separate}
\\ \nonumber &=& \left( f^T_{h_c} \right)^2 \epsilon_{\alpha\beta\mu\nu} \epsilon^\mu p^\nu \epsilon_{\alpha^\prime\beta^\prime\mu^\prime\nu^\prime} \epsilon^{*\mu^\prime} p^{\nu^\prime}
\\ \nonumber &=& - \left( f^T_{h_c} \right)^2 ~ p^2 ~ \left( g_{\alpha \alpha^\prime} g_{\beta \beta^\prime} - g_{\alpha \beta^\prime} g_{\beta \alpha^\prime} \right) + \cdots \, ,
\end{eqnarray}
since the correlation function of $J/\psi$ dose not contain the above coefficient. It is not so easy to isolate $J/\psi$ from $J_{\alpha\beta}$ at the hadron level. Instead, we can investigate its partner current
\begin{equation}
\tilde J_{\alpha\beta} = \epsilon_{\alpha\beta\gamma\delta} \times J^{\gamma\delta} \, ,
\end{equation}
which couples to $J/\psi$ and $h_c$ just in the opposite ways:
\begin{eqnarray}
\langle 0 | \tilde J_{\alpha\beta} | J/\psi (\epsilon,p) \rangle &=& i \tilde f^T_{J/\psi} \epsilon_{\alpha\beta\mu\nu} \epsilon^\mu p^\nu \, ,
\\ \langle 0 | \tilde J_{\alpha\beta} | h_c (\epsilon,p) \rangle &=& i \tilde f^T_{h_c} (p_\alpha\epsilon_\beta - p_\beta\epsilon_\alpha) \, .
\end{eqnarray}
Accordingly, we can use the two currents $J_{\alpha\beta}$ and $\tilde J_{\alpha\beta}$ to study and well separate $J/\psi$ and $h_c$.

We apply the above process to generally investigate the current $J^{\alpha_1\cdots\alpha_N,\beta_1\cdots\beta_N}$, which has $2 N = 2 J$ Lorentz indices with certain symmetries, {\it e.g.}, the spin-2 current $J^{\alpha_1\alpha_2,\beta_1\beta_2}$ has four Lorentz indices, satisfying
\begin{equation}
J^{\alpha_1\alpha_2,\beta_1\beta_2} = - J^{\beta_1\alpha_2,\alpha_1\beta_2} = - J^{\alpha_1\beta_2,\beta_1\alpha_2} = J^{\alpha_2\alpha_1,\beta_2\beta_1} \, .
\end{equation}
Its coupling can be written as:
\begin{equation}
\langle 0 | J^{\alpha_1\cdots\alpha_N,\beta_1\cdots\beta_N} | X \rangle = i f_X \mathcal{S}[ \epsilon^{\alpha_i \beta_i \mu_i \nu_i} p_{\nu_i} ]^N \epsilon_{\mu_1 \cdots \mu_N} \, ,
\end{equation}
where $X$ is the corresponding state having the same parity as $J^{i_1 \cdots i_N,j_1 \cdots j_N}$ ($i_1 \cdots j_N=1,2,3$); $\mathcal{S}$ denotes symmetrization and subtracting trace terms in the two sets $\{\alpha_1 \cdots \alpha_N\}$ and $\{\beta_1 \cdots \beta_N\}$ simultaneously, with
\begin{equation}
[ \cdots ]^N = \epsilon^{\alpha_1 \beta_1 \mu_1 \nu_1} p_{\nu_1} \cdots \epsilon^{\alpha_N \beta_N \mu_N \nu_N} p_{\nu_N} \, .
\end{equation}
Note that the current $J^{\alpha_1\cdots\alpha_N,\beta_1\cdots\beta_N}$ can also couple to the other state $X^\prime$ having the parity opposite to $X$, but this state $X^\prime$ can not be easily isolated at the hadron level, so we do not consider it in the present study.

\subsection{Two-gluon glueball currents}

In this subsection we use the gluon field strength tensor $G^a_{\mu\nu}$ to construct two-gluon glueball currents, with $a$ the color index and $\mu,\nu$ the Lorentz indices. We also need $\tilde G^a_{\mu\nu} = G^{a,\rho\sigma} \times \epsilon_{\mu\nu\rho\sigma}/2$ to denote the dual gluon field strength tensor, and $f^{abc}$ to denote the totally antisymmetric $SU(3)_C$ structure constant. In the present study we only consider local glueball currents without explicit derivatives, although $G^a_{\mu\nu}$ and $\tilde G^a_{\mu\nu}$ contain covariant derivatives inside themselves.

In Ref.~\cite{Jaffe:1985qp} the authors use the chromoelectric and chromomagnetic fields ($i,j=1,2,3$),
\begin{equation}
E_i = G_{i0}
~~~~ {\rm and} ~~~~
B_i = -{1\over2} \epsilon_{ijk} G^{jk} \, ,
\end{equation}
to write down all the non-relativistic low-dimension two-gluon glueball operators:
\begin{eqnarray}
\nonumber && 0^{++} ~~~~~ \vec E_a^2 \pm \vec B_a^2 \, ,
\\ \nonumber && 0^{-+} ~~~~~ \vec E_a \cdot \vec B_a \, ,
\\ && 1^{-+} ~~~~~ \vec E_a \times \vec B_a \, ,
\\ \nonumber && 2^{++} ~~~~~ \mathcal{S}^\prime[ E_a^i E_a^j \pm B_a^i B_a^j ] \, ,
\\ \nonumber && 2^{-+} ~~~~~ \mathcal{S}^\prime[ E_a^i B_a^j - B_a^i E_a^j ] \, ,
\end{eqnarray}
where $\mathcal{S}^\prime$ denotes symmetrization and subtracting trace terms in the set $\{ij\}$.

We construct their corresponding relativistic currents in order to perform QCD sum rule analyses:
\begin{eqnarray}
J_0 &=& g_s^2 G_a^{\mu\nu} G^{a}_{\mu\nu} \, ,
\label{def:J0}
\\[1mm]
\tilde J_0 &=& g_s^2 G_a^{\mu\nu} \tilde G^{a}_{\mu\nu} \, ,
\label{def:J0t}
\\[1mm]
J_1^{\alpha\beta} &=& g_s^2 G_a^{\alpha\mu} \tilde G^{a,\beta}_{\mu} - \{ \alpha \leftrightarrow \beta \} \, ,
\label{def:J1}
\\[1mm]
J_2^{\alpha_1\alpha_2,\beta_1\beta_2} &=& \mathcal{S}[ g_s^2 G_a^{\alpha_1\beta_1} G^{a,\alpha_2\beta_2} ] \, ,
\\[1mm]
\tilde J_2^{\alpha_1\alpha_2,\beta_1\beta_2} &=& \mathcal{S}[ g_s^2 G_a^{\alpha_1\beta_1} \tilde G^{a,\alpha_2\beta_2} ] \, .
\end{eqnarray}
We shall explicitly prove in Appendix~\ref{app:spin1} that the third current $J_1^{\alpha\beta}$ vanishes, suggesting that the ``ground-state'' two-gluon glueball of $J^{PC} = 1^{-+}$ does not exist within the relativistic framework.

The former two currents $J_0$ of $J^{PC} = 0^{++}$ and $\tilde J_0$ of $J^{PC} = 0^{-+}$ couple to the $J^{PC} = 0^{++}$ and $0^{-+}$ two-gluon glueballs $|{\rm GG};{J^{PC}}\rangle$, respectively:
\begin{eqnarray}
\langle 0 | J_0 | {\rm GG};{0^{++}} \rangle &=& f_{0^{++}} \, ,
\\
\langle 0 | \tilde J_0 | {\rm GG};{0^{-+}} \rangle &=& f_{0^{-+}} \, ,
\end{eqnarray}
where $f_{0^{++}}$ and $f_{0^{-+}}$ are decay constants. Besides, the current $J_0$ has a partner,
\begin{equation}
J_0^\prime = g_s^2 \tilde G_a^{\mu\nu} \tilde G^{a}_{\mu\nu} \, ,
\end{equation}
whose sum rule result is the same as that of $J_0$.

The latter two currents $J_2^{\alpha_1\alpha_2,\beta_1\beta_2}$ and $\tilde J_2^{\alpha_1\alpha_2,\beta_1\beta_2}$ couple to the $J^{PC} = 2^{++}$ and $2^{-+}$ glueballs through:
\begin{eqnarray}
\langle 0 | J_2^{\cdots} | {\rm GG};{2^{++}} \rangle &=& i f_{2^{++}} \mathcal{S}[ \epsilon^{\alpha_i \beta_i \mu_i \nu_i} p_{\nu_i} ]^2 \epsilon_{\mu_1\mu_2} \, ,
\\ \langle 0 | \tilde J_2^{\cdots} | {\rm GG};{2^{-+}} \rangle &=& i f_{2^{-+}} \mathcal{S}[ \epsilon^{\alpha_i \beta_i \mu_i \nu_i} p_{\nu_i} ]^2 \epsilon_{\mu_1\mu_2} \, .
\end{eqnarray}
The current $J_2^{\alpha_1\alpha_2,\beta_1\beta_2}$ also has a partner,
\begin{equation}
J_2^{\prime\alpha_1\alpha_2,\beta_1\beta_2} = \mathcal{S}[ g_s^2 \tilde G_a^{\alpha_1\beta_1} \tilde G^{a,\alpha_2\beta_2} ] \, .
\end{equation}
whose sum rule result is the same as that of $J_2^{\alpha_1\alpha_2,\beta_1\beta_2}$.

\subsection{Three-gluon glueball currents of $C=+$}

In this subsection we use $G^a_{\mu\nu}$ and $\tilde G^a_{\mu\nu}$ to construct three-gluon glueball currents of $C=+$. Some of their corresponding non-relativistic operators have been constructed in Ref.~\cite{Jaffe:1985qp}:
\begin{eqnarray}
\nonumber && 0^{++} ~~~~~ f^{abc} ( \vec E_a \times \vec E_b ) \cdot \vec B_c \, ,
\\ \nonumber && 0^{-+} ~~~~~ f^{abc} ( \vec E_a \times \vec E_b ) \cdot \vec E_c \, ,
\\ && 1^{++} ~~~~~ f^{abc} ( \vec B_a \cdot \vec E_b )~\vec E_c \, ,
\\ \nonumber && 1^{-+} ~~~~~ f^{abc} ( \vec B_a \cdot \vec E_b )~\vec B_c \, ,
\\ \nonumber && 2^{++} ~~~~~ f^{abc} \mathcal{S}^\prime[ ( \vec B_a \times \vec B_b )^i B_c^j ] + \cdots \, ,
\\ \nonumber && 2^{-+} ~~~~~ f^{abc} \mathcal{S}^\prime[ ( \vec E_a \times \vec E_b )^i E_c^j ] + \cdots \, .
\end{eqnarray}
We further construct their corresponding relativistic currents as follows:
\begin{eqnarray}
\eta_0 &=& f^{abc} g_s^3 G_a^{\mu\nu} G_{b,\nu\rho} G_{c,\mu}^{\rho} \, ,
\label{def:3G1}
\\[1mm]
\tilde \eta_0 &=& f^{abc} g_s^3 \tilde G_a^{\mu\nu} \tilde G_{b,\nu\rho} \tilde G_{c,\mu}^{\rho} \, ,
\label{def:3G2}
\\[1mm]
\eta_1^{\alpha\beta} &=& f^{abc} g_s^3 \tilde G_a^{\mu\nu} G_{b,\mu\nu} \tilde G_{c}^{\alpha\beta} \, ,
\label{def:eta1}
\\[1mm]
\tilde \eta_1^{\alpha\beta} &=& f^{abc} g_s^3 \tilde G_a^{\mu\nu} G_{b,\mu\nu} G_{c}^{\alpha\beta} \, ,
\label{def:eta1t}
\\[1mm] \nonumber
\eta_2^{\alpha_1\alpha_2,\beta_1\beta_2} &=& f^{abc} \mathcal{S}[ g_s^3 G_a^{\alpha_1\beta_1} G_b^{\alpha_2\mu} G_{c,\mu}^{\beta_2} - \{ \alpha_2 \leftrightarrow \beta_2 \} ] \, ,
\\
\\[1mm] \nonumber
\tilde \eta_2^{\alpha_1\alpha_2,\beta_1\beta_2} &=& f^{abc} \mathcal{S}[ g_s^3 \tilde G_a^{\alpha_1\beta_1} \tilde G_b^{\alpha_2\mu} \tilde G_{c,\mu}^{\beta_2} - \{ \alpha_2 \leftrightarrow \beta_2 \} ] \, .
\\
\end{eqnarray}
We shall explicitly prove in Appendix~\ref{app:spin1} that the third and fourth currents $\eta_1^{\alpha\beta}$ and $\tilde \eta_1^{\alpha\beta}$ both vanish, suggesting that the ``ground-state'' three-gluon glueballs of $J^{PC} = 1^{++}$ and $1^{-+}$ do not exist within the relativistic framework.

The former two currents $\eta_0$ of $J^{PC} = 0^{++}$ and $\tilde \eta_0$ of $J^{PC} = 0^{-+}$ couple to the $J^{PC} = 0^{++}$ and $0^{-+}$ three-gluon glueballs $|{\rm GGG};{J^{PC}}\rangle$, respectively:
\begin{eqnarray}
\langle 0 | \eta_0 | {\rm GGG};{0^{++}} \rangle &=& f^\prime_{0^{++}} \, ,
\\
\langle 0 | \tilde \eta_0 | {\rm GGG};{0^{-+}} \rangle &=& f^\prime_{0^{-+}} \, .
\end{eqnarray}
The latter two currents $\eta_2^{\alpha_1\alpha_2,\beta_1\beta_2}$ and $\tilde \eta_2^{\alpha_1\alpha_2,\beta_1\beta_2}$ couple to the $J^{PC} = 2^{++}$ and $2^{-+}$ glueballs through:
\begin{eqnarray}
\langle 0 | \eta_2^{\cdots} | {\rm GGG};{2^{++}} \rangle &=& i f^\prime_{2^{++}} \mathcal{S}[ \epsilon^{\alpha_i \beta_i \mu_i \nu_i} p_{\nu_i} ]^2 \epsilon_{\mu_1\mu_2} ,
\\ \langle 0 | \tilde \eta_2^{\cdots} | {\rm GGG};{2^{-+}} \rangle &=& i f^\prime_{2^{-+}} \mathcal{S}[ \epsilon^{\alpha_i \beta_i \mu_i \nu_i} p_{\nu_i} ]^2 \epsilon_{\mu_1\mu_2} .
\end{eqnarray}

%
\section{QCD sum rule analyses}
\label{sec:sumrule}
%

In this section we use the two-gluon glueball currents $J_0$, $\tilde J_0$, $J_2^{\alpha_1\alpha_2,\beta_1\beta_2}$, and $\tilde J_2^{\alpha_1\alpha_2,\beta_1\beta_2}$ as well as the three-gluon glueball currents $\eta_0$, $\tilde \eta_0$, $\eta_2^{\alpha_1\alpha_2,\beta_1\beta_2}$, and $\tilde \eta_2^{\alpha_1\alpha_2,\beta_1\beta_2}$ to perform QCD sum rule analyses. This method has been widely applied in the field of hadron phenomenology~\cite{Shifman:1978bx,Reinders:1984sr} to study various exotic hadrons~\cite{Chen:2007xr,Chen:2016otp,Chen:2016jxd}. Especially, all the above spin-2 currents have four Lorentz indices with certain symmetries, so that they couple to both positive- and negative-parity glueballs simultaneously. We refer to Ref.~\cite{Chen:2021cjr} for detailed discussions.

We take the current $\tilde J_0$ defined in Eq.~(\ref{def:J0t}) as an example, and calculate its two-point correlation function
%
\begin{equation}
\Pi(q^2) \equiv i \int d^4x e^{iqx} \langle 0 | {\bf T}[\tilde J_0(x) \tilde J_0^\dagger(0)] | 0 \rangle \, ,
\label{def:pi}
\end{equation}
%
separately at hadron and quark-gluon levels.

At the hadron level we express Eq.~(\ref{def:pi}) using the dispersion relation as
%
\begin{equation}
\Pi(q^2) = \int^\infty_{0}\frac{\rho(s)}{s-q^2-i\varepsilon}ds \, ,
\label{def:rho}
\end{equation}
%
with $\rho(s) = {\rm Im}\Pi(s)/\pi$ the spectral density. It is parameterized using one
pole dominance for the ground state $X$ as well as the continuum contribution,
%
\begin{eqnarray}
\nonumber \rho(s) &\equiv& \sum_n\delta(s-M^2_n) \langle 0| \tilde J_0 | n\rangle \langle n| \tilde J_0^{\dagger} |0 \rangle
\\ &=& f^2_X \delta(s-M^2_X) + \rm{continuum} \, .
\label{eq:rho}
\end{eqnarray}
%

At the quark-gluon level we insert Eq.~(\ref{def:J0t}) into Eq.~(\ref{def:pi}), and calculate it using the method of operator product expansion (OPE). After performing the Borel transformation to Eq.~(\ref{def:rho}) at both hadron and quark-gluon levels, we approximate the continuum using the spectral density above a threshold value $s_0$, and obtain
%
\begin{equation}
\Pi(s_0, M_B^2) \equiv f^2_X e^{-M_X^2/M_B^2} = \int^{s_0}_{0} e^{-s/M_B^2}\rho(s)ds \, .
\label{eq:fin}
\end{equation}
%
This equation can be used to further calculate the mass of $X$ through
%
\begin{equation}
M^2_X(s_0, M_B) = \frac{\int^{s_0}_{0} e^{-s/M_B^2}s\rho(s)ds}{\int^{s_0}_{0} e^{-s/M_B^2}\rho(s)ds} \, .
\label{eq:LSR}
\end{equation}
%

\begin{figure}[hbtp]
\begin{center}
\subfigure[($a$)]{
\scalebox{0.35}{\includegraphics{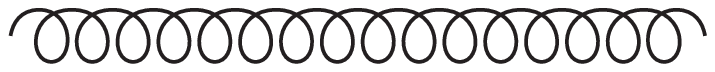}}}
\subfigure[($b$)]{
\scalebox{0.35}{\includegraphics{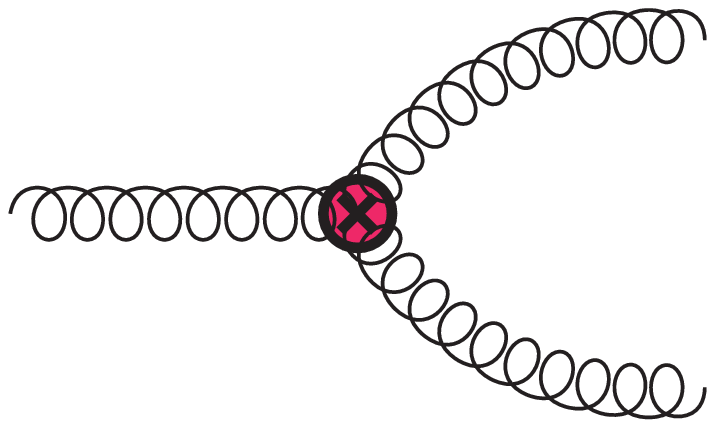}}}
\end{center}
\caption{The gluon field strength tensor $G^a_{\mu\nu} = \partial_\mu A_\nu^a - \partial_\nu A_\mu^a + g_s f^{abc} A_{b,\mu} A_{c,\nu}$, naturally separated into two parts ($a$) and ($b$).}
\label{fig:gluon}
\end{figure}

\begin{figure*}[hbtp]
\begin{center}
\subfigure[($a$)]{
\scalebox{0.2}{\includegraphics{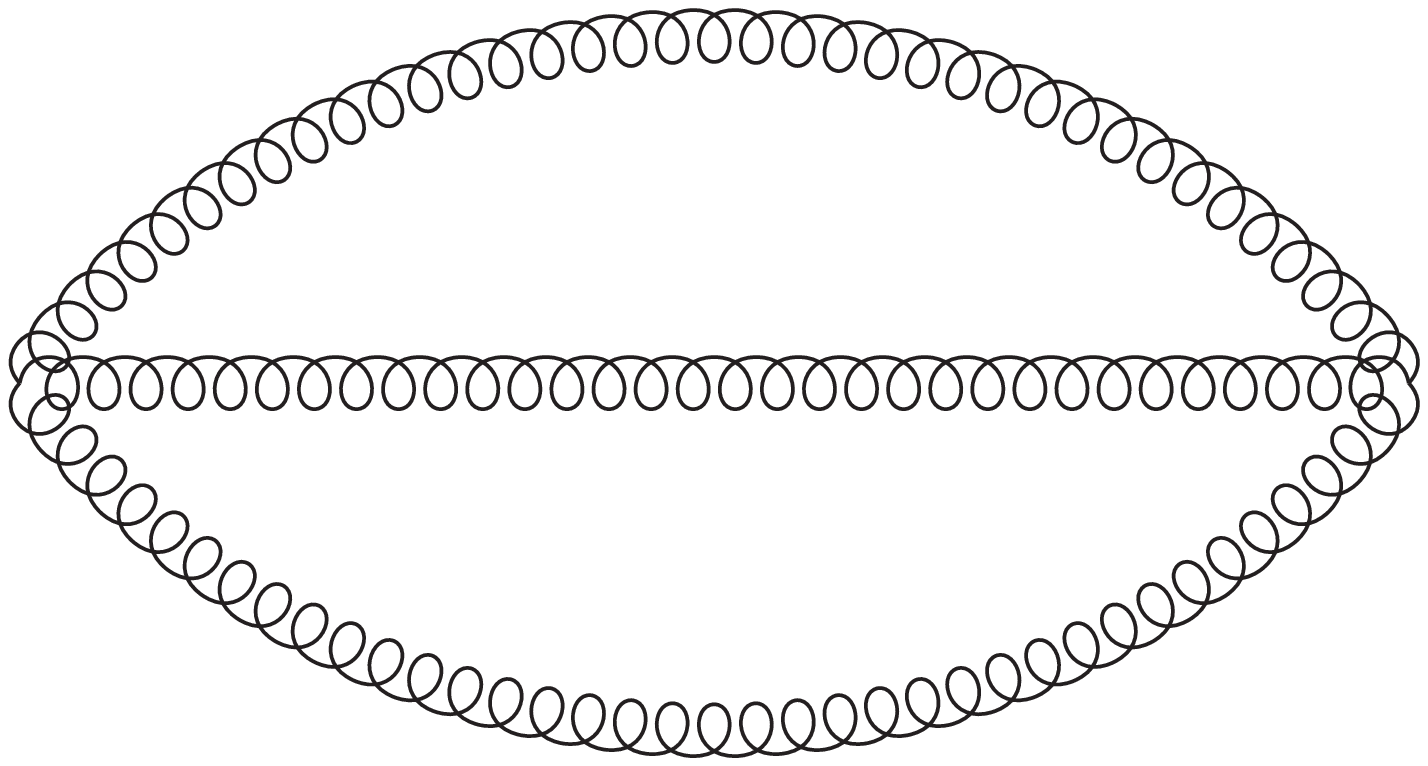}}}
\\[5mm]
\subfigure[($b{\rm-}1$)]{
\scalebox{0.2}{\includegraphics{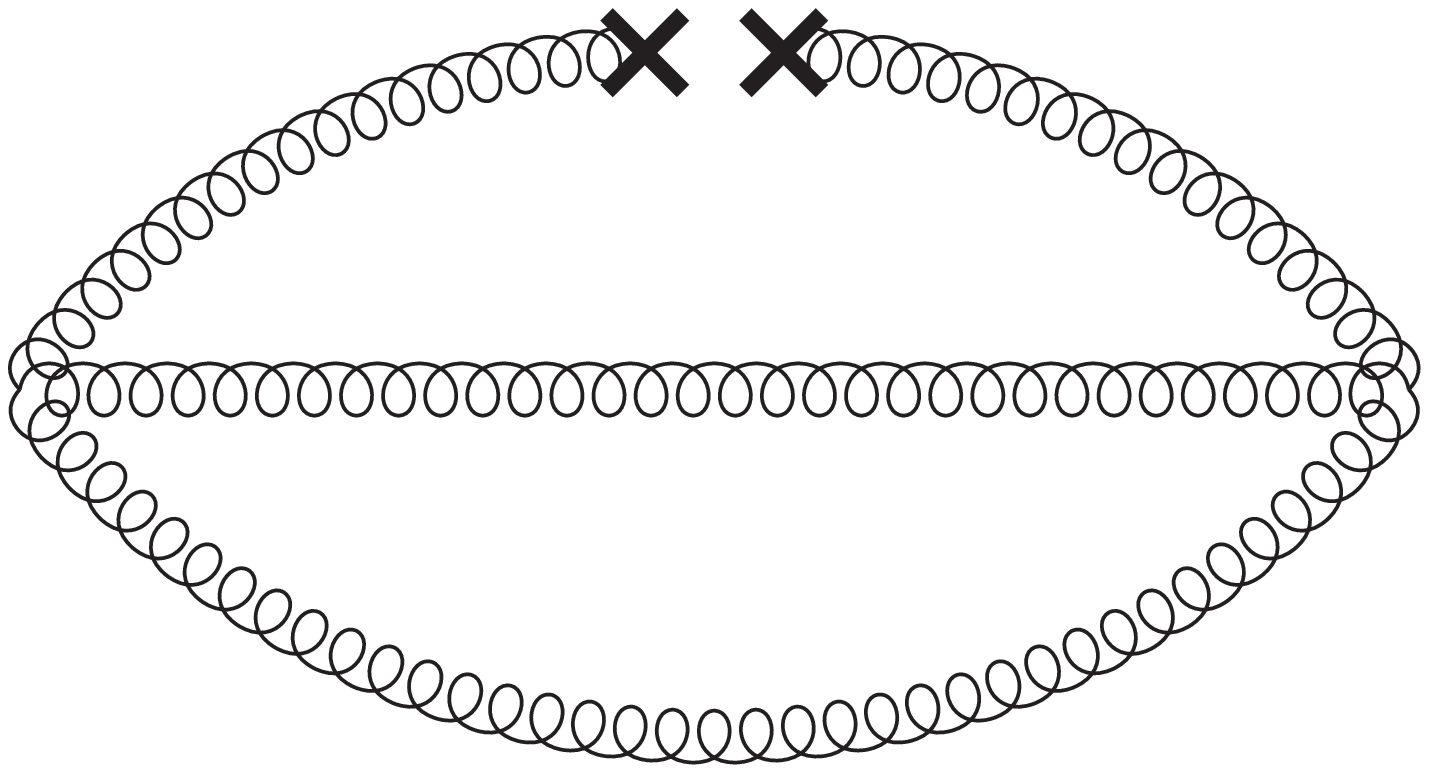}}}~~~~~~~~~~~~~~~~~~~~~~~~~~~~~~~~~~~~
\subfigure[($b{\rm-}2$)]{
\scalebox{0.2}{\includegraphics{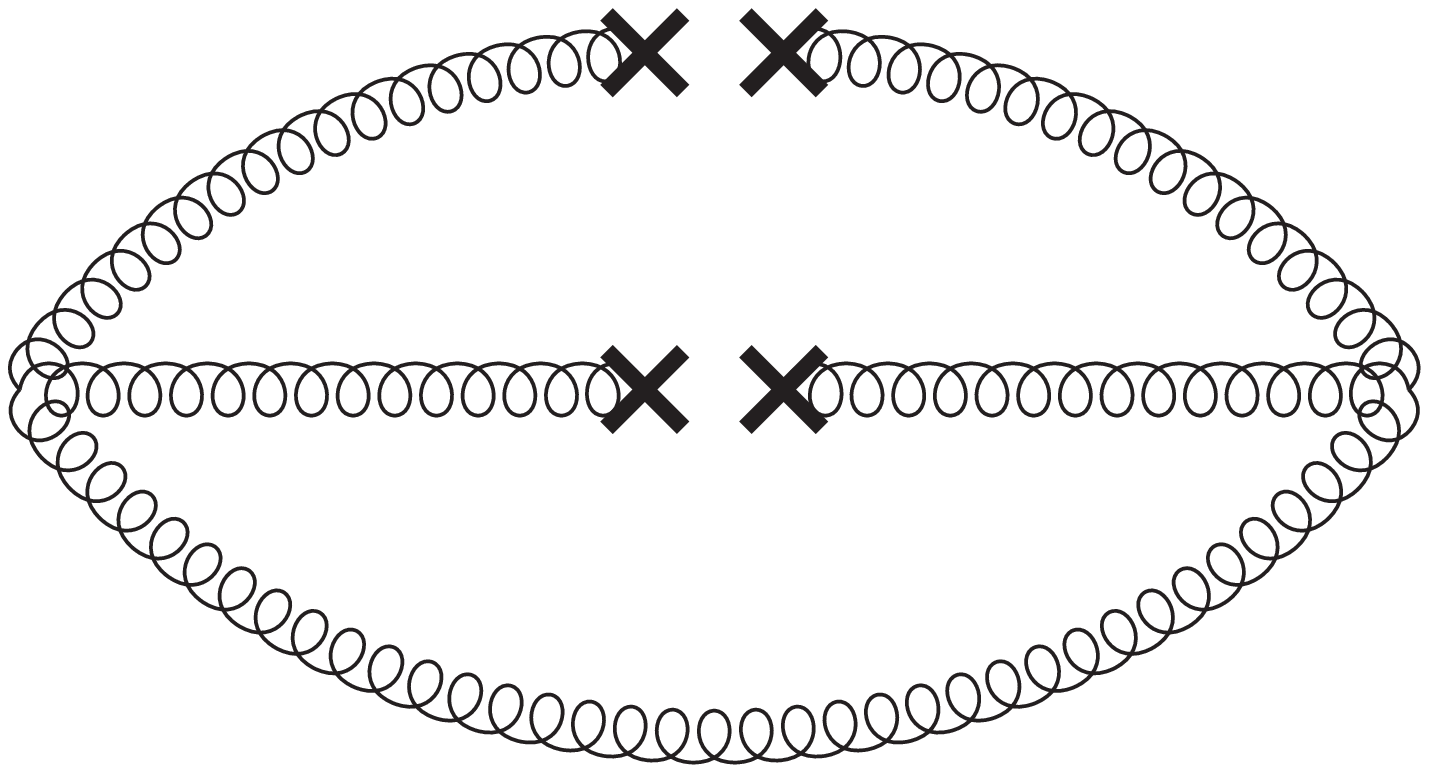}}}
\\[5mm]
\subfigure[($c{\rm-}1$)]{
\scalebox{0.2}{\includegraphics{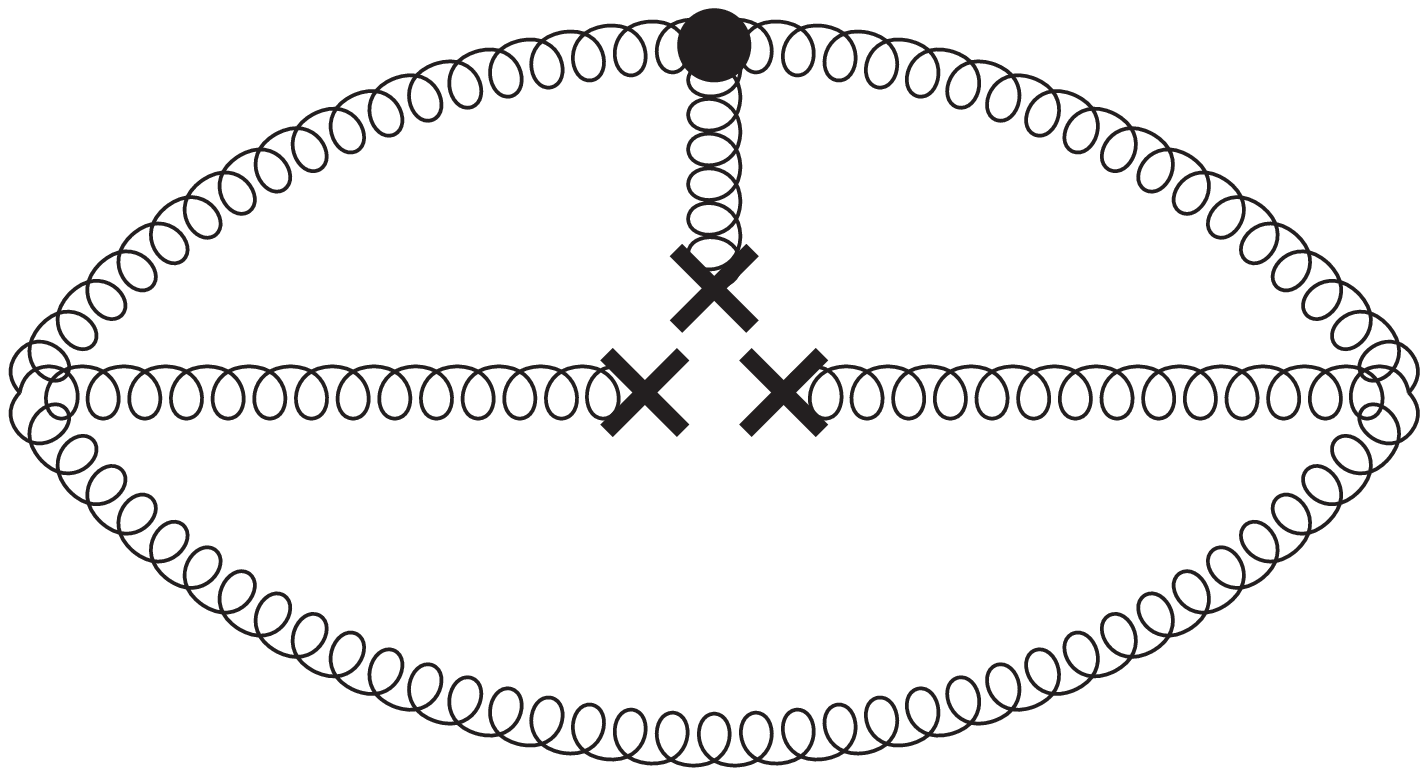}}}~~~
\subfigure[($c{\rm-}2$)]{
\scalebox{0.2}{\includegraphics{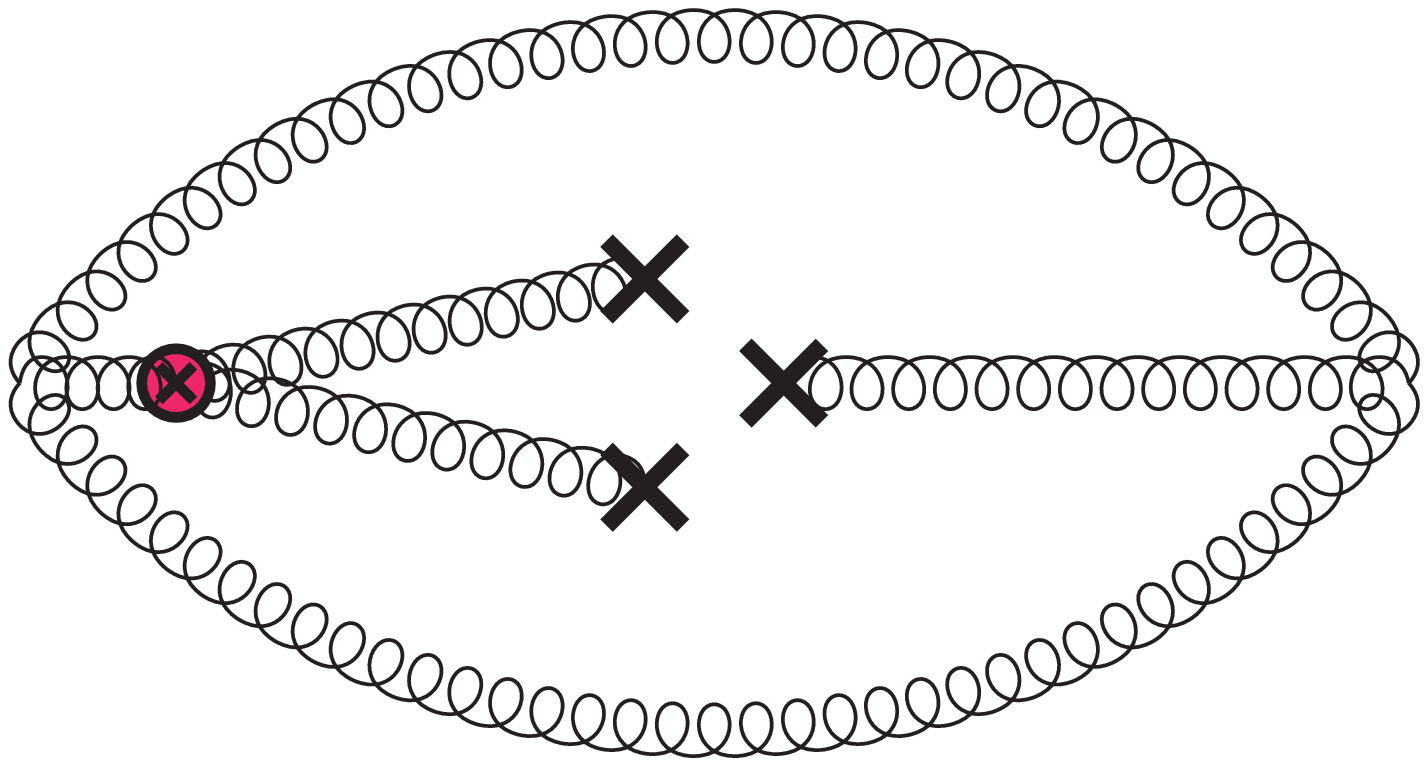}}}~~~
\subfigure[($c{\rm-}3$)]{
\scalebox{0.2}{\includegraphics{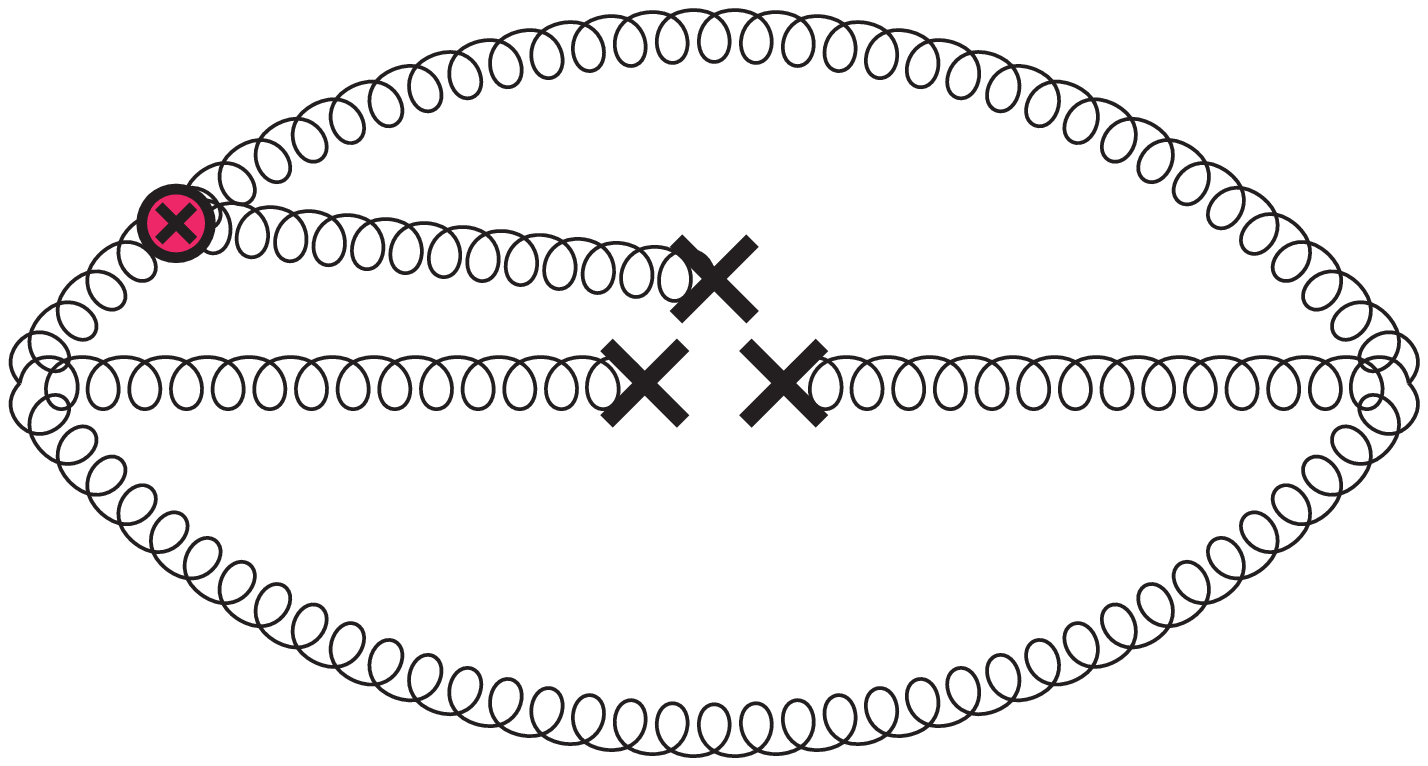}}}~~~
\subfigure[($c{\rm-}4$)]{
\scalebox{0.2}{\includegraphics{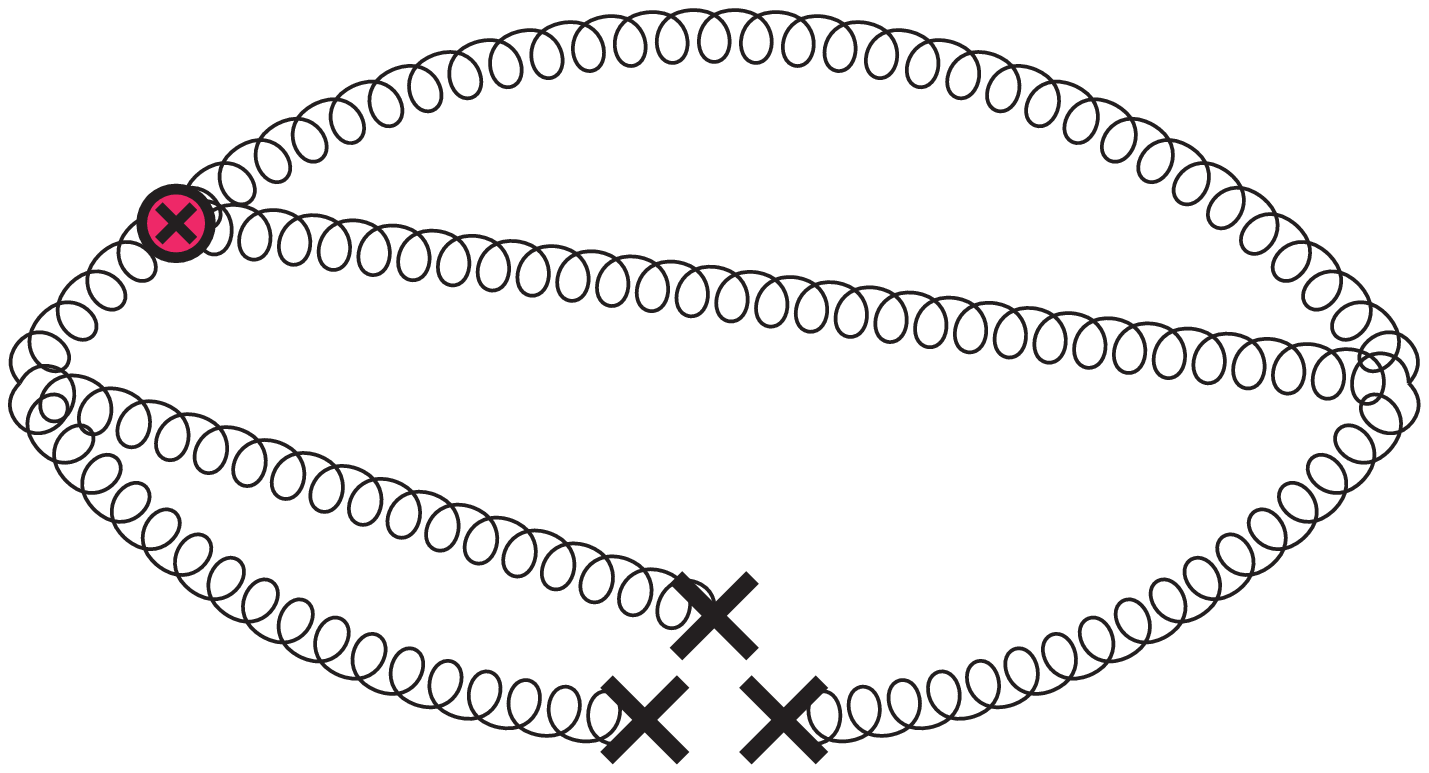}}}~~~
\subfigure[($c{\rm-}5$)]{
\scalebox{0.2}{\includegraphics{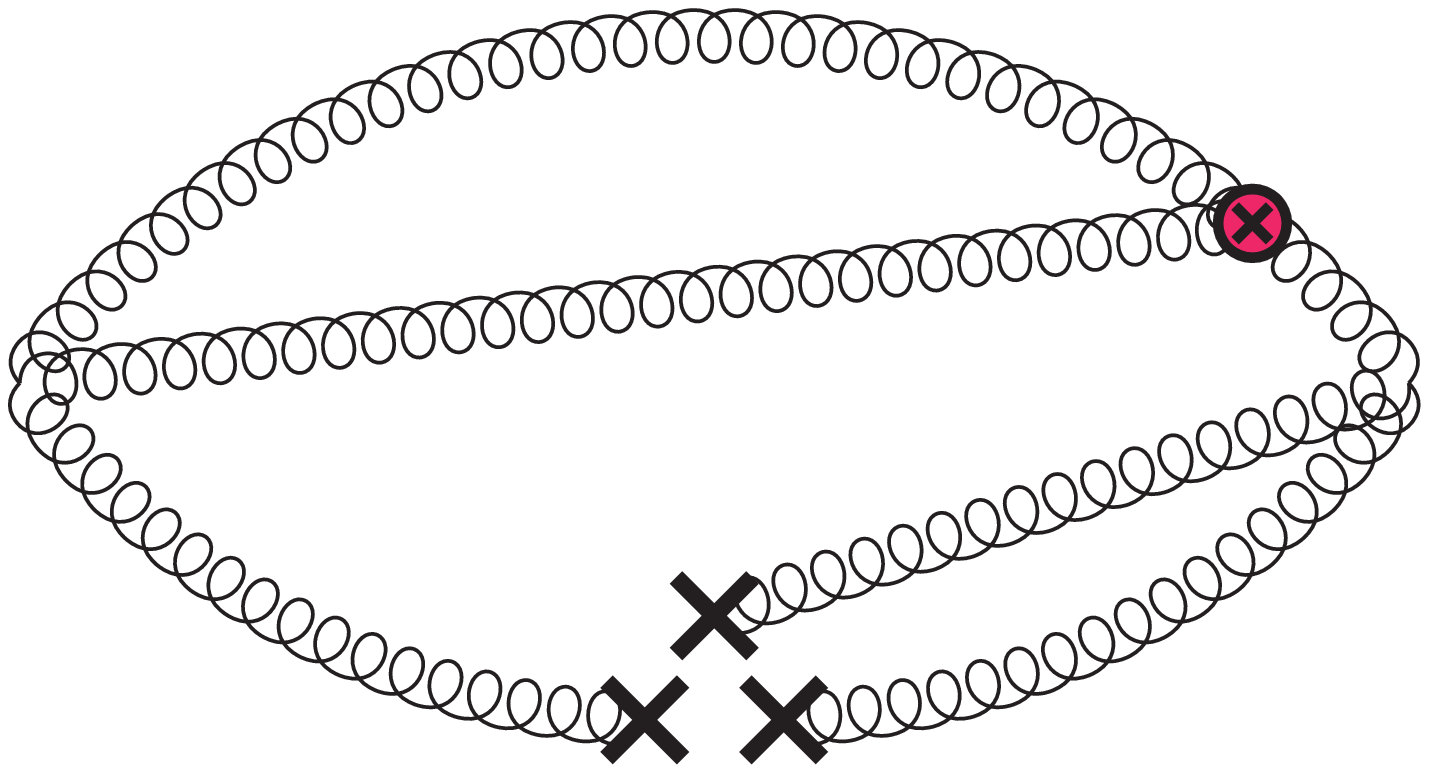}}}
\\[5mm]
\subfigure[($d$)]{
\scalebox{0.2}{\includegraphics{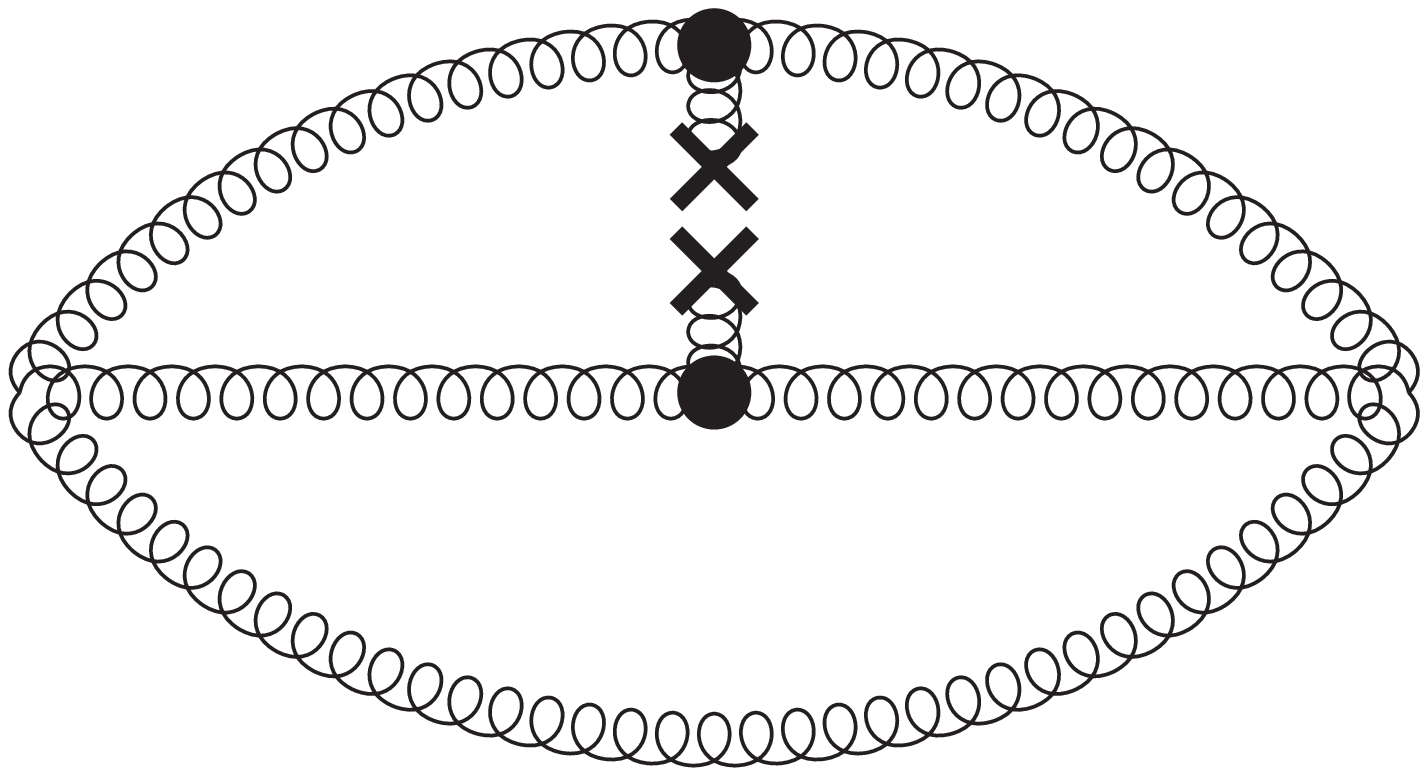}}}
\end{center}
\caption{Feynman diagrams for three-gluon glueball currents, including the perturbative term, the two-gluon condensate $\langle g_s^2 GG\rangle$, the three-gluon condensate $\langle g_s^3 G^3 \rangle$, and the $D=8$ condensate $\langle g_s^2 GG\rangle^2$. The diagrams $(a)$ and $(b-i)$ are proportional to $\alpha_s^3 \times g_s^0$, the diagrams $(c-i)$ are proportional to $\alpha_s^3 \times g_s^1$, and the diagram $(d)$ is proportional to $\alpha_s^3 \times g_s^2$.}
\label{fig:feynman}
\end{figure*}

Since the gluon field strength tensor $G^a_{\mu\nu}$ is defined as
\begin{equation}
G^a_{\mu\nu} = \partial_\mu A_\nu^a - \partial_\nu A_\mu^a + g_s f^{abc} A_{b,\mu} A_{c,\nu} \, ,
\end{equation}
it can be naturally separated into two parts. As shown in Fig.~\ref{fig:gluon}, we depict the former two terms using the single-gluon-line, and the latter one term using the double-gluon-line with a red vertex (see also the diagram Fig.~\ref{fig:feynman}$(c-3)$). Here $A_\mu^a$ is the gluon field, whose propagator is~\cite{Govaerts:1984bk}:
\begin{eqnarray}
\langle 0 | {\bf T}[ A_\mu^a(x) A_\nu^b(y) ] | 0 \rangle &=& {\delta^{ab} g_{\mu\nu} \over 4 \pi^2 (x-y)^2}
\\ \nonumber &+& {g_s \ln(-(x-y)^2) \over 8 \pi^2} f^{abc} G_{c,\mu\nu}(0)
\\ \nonumber &-& {g_s g_{\mu\nu} x^\alpha y^\beta \over 8 \pi^2 (x-y)^2} f^{abc} G_{c,\alpha\beta}(0) \, .
\label{eq:gluon}
\end{eqnarray}
We work in the fixed-point gauge so that
\begin{equation}
A_\mu^a(x) \approx - {1\over2} x^\nu G^a_{\mu\nu}(0) \, .
\end{equation}

In the present study we consider the Feynman diagrams depicted in Fig.~\ref{fig:feynman} (for three-gluon glueballs), and calculate OPEs up to the dimension eight ($D=8$) condensates. We take into account the perturbative term, the two-gluon condensate $\langle g_s^2 GG\rangle$, the three-gluon condensate $\langle g_s^3 G^3 \rangle$, and the $D=8$ condensate $\langle g_s^2 GG\rangle^2$:
\begin{eqnarray}
&& \Pi_{|{\rm GG};0^{++}\rangle}(s_0, M_B^2)
\label{eq:GGope0pp}
\\ \nonumber &=& \int_0^{s_0} \Big( 32 \alpha_s^2 s^2 + 60 \alpha_s^2 \langle g_s^2 GG \rangle \Big) e^{-s/M_B^2} ds
\\ \nonumber && ~~~~~~~~~~~~~~~~~~~~~~~~~~~~~~~ + 24 \pi \alpha_s \langle g_s^3 G^3 \rangle  ,
\\[2mm]
&& \Pi_{|{\rm GG};2^{++}\rangle}(s_0, M_B^2)
\label{eq:GGope2pp}
\\ \nonumber &=& \int_0^{s_0} \Big( {2 \alpha_s^2 \over 15} s^2 - {5 \alpha_s^2 \langle g_s^2 GG \rangle \over 24} \Big) e^{-s/M_B^2} ds
\\ \nonumber && ~~~~~~~~~~~~~~~~~~~~~~~~~~~~~~~~ + {\pi \alpha_s \langle g_s^3 G^3 \rangle \over 3}  ,
\\[2mm]
&& \Pi_{|{\rm GG};0^{-+}\rangle}(s_0, M_B^2)
\label{eq:GGope0mp}
\\ \nonumber &=& \int_0^{s_0} 32 \alpha_s^2 s^2 e^{-s/M_B^2} ds - 40 \pi \alpha_s \langle g_s^3 G^3 \rangle  ,
\\[2mm]
&& \Pi_{|{\rm GG};2^{-+}\rangle}(s_0, M_B^2)
\label{eq:GGope2mp}
\\ \nonumber &=& \int_0^{s_0} \Big( {2 \alpha_s^2 \over 5} s^2 + { \alpha_s^2 \langle g_s^2 GG \rangle \over 12} \Big) e^{-s/M_B^2} ds
\\ \nonumber && ~~~~~~~~~~~~~~~~~~~~~~~~~~~~~ - {\pi \alpha_s \langle g_s^3 G^3 \rangle \over 2}  ,
\\[2mm]
&& \Pi_{|{\rm GGG};0^{++}\rangle}(s_0, M_B^2)
\label{eq:GGGope0pp}
\\ \nonumber &=& \int_0^{s_0} \Big( {3 \alpha_s^3 \over 10 \pi} s^4 + {135 \alpha_s^3 \langle g_s^2 GG \rangle \over 32 \pi} s^2
\\ \nonumber && ~~~~~~~~~~~~~~~~~~~~ - {81 \alpha_s^2 \langle g_s^3 G^3 \rangle \over 2} s \Big) e^{-s/M_B^2} ds  ,
\\[2mm]
&& \Pi_{|{\rm GGG};2^{++}\rangle}(s_0, M_B^2)
\label{eq:GGGope2pp}
\\ \nonumber &=& \int_0^{s_0} \Big( {2 \alpha_s^3 \over 315 \pi} s^4 + {\alpha_s^2 \langle g_s^2 GG \rangle \over 15} s^2
\\ \nonumber && ~~~~~~ + {53 \alpha_s^3 \langle g_s^2 GG \rangle \over 320 \pi} s^2 + {\alpha_s^2 \langle g_s^3 G^3 \rangle \over 3} s \Big) e^{-s/M_B^2} ds  ,
\\[2mm]
&& \Pi_{|{\rm GGG};0^{-+}\rangle}(s_0, M_B^2)
\label{eq:GGGope0mp}
\\ \nonumber &=& \int_0^{s_0} \Big( {3 \alpha_s^3 \over 10 \pi} s^4 + {135 \alpha_s^3 \langle g_s^2 GG \rangle \over 32 \pi} s^2
\\ \nonumber && ~~~~~~~~~~~~~~~~~~~~ + {27 \alpha_s^2 \langle g_s^3 G^3 \rangle \over 2} s \Big) e^{-s/M_B^2} ds  ,
\\[2mm]
&& \Pi_{|{\rm GGG};2^{-+}\rangle}(s_0, M_B^2)
\label{eq:GGGope2mp}
\\ \nonumber &=& \int_0^{s_0} \Big( {2 \alpha_s^3 \over 315 \pi} s^4 - {\alpha_s^2 \langle g_s^2 GG \rangle \over 15} s^2
\\ \nonumber && ~~~~~ + {57 \alpha_s^3 \langle g_s^2 GG \rangle \over 320 \pi} s^2 + {5 \alpha_s^2 \langle g_s^3 G^3 \rangle \over 4} s \Big) e^{-s/M_B^2} ds  .
\end{eqnarray}
In the calculations we have considered all the diagrams proportional to $\alpha_s^n \times g_s^0$ and $\alpha_s^n \times g_s^1$ ($n=2$ for two-gluon glueballs and $n=3$ for three-gluon glueballs); however, there are so many diagrams proportional to $\alpha_s^n \times g_s^2$, so we have only taken into account one of them. Especially, we find all the $D=8$ terms proportional to $\langle g_s^2 GG \rangle^2$ vanish, so the convergence of the above OPE series are quite good.

In Ref.~\cite{Latorre:1987wt} the authors studied $J^{PC} = 0^{++}$ three-gluon glueballs using the current $\eta_0$ defined in Eq.~(\ref{def:3G1}), where they calculated the Feynman diagrams depicted in Figs.~\ref{fig:feynman}($a,b-i,c-1,c-2$). In Ref.~\cite{Hao:2005hu} the authors studied $J^{PC} = 0^{-+}$ three-gluon glueballs using the current $\tilde \eta_0$ defined in Eq.~(\ref{def:3G2}), where they calculated the diagrams depicted in Figs.~\ref{fig:feynman}($a,b-i,c-i$). Their calculations are done (mainly) by hand. In the present study we use the software {\it Mathematica} with the package {\it FeynCalc}, and we can obtain exactly the same results for these diagrams. In Refs.~\cite{Novikov:1981xi,Bagan:1990sy,Forkel:2003mk} the authors studied $J^{PC} = 0^{++}$ and $0^{-+}$ two-gluon glueballs using the currents $J_0$ and $\tilde J_0$ defined in Eqs.~(\ref{def:J0}) and (\ref{def:J0t}), where they calculated more diagrams than those calculated in the present study. However, such calculations are too complicated to be applied to three-gluon glueballs, and we still calculate similar diagrams as those depicted in Fig.~\ref{fig:feynman} for two-gluon glueballs to make the present study unified as a whole.

For completeness, we also investigate the following three-gluon glueball currents of $C=-$:
\begin{eqnarray}
\xi_1^{\alpha\beta} &=& d^{abc} g_s^3 G_a^{\mu\nu} G_{b,\mu\nu} G_{c}^{\alpha\beta} \, ,
\\ \tilde \xi_1^{\alpha\beta} &=& d^{abc} g_s^3 G_a^{\mu\nu} G_{b,\mu\nu} \tilde G_{c}^{\alpha\beta} \, ,
\\ \nonumber \xi_2^{\alpha_1\alpha_2,\beta_1\beta_2} &=& d^{abc} \mathcal{S}[ g_s^3 \tilde G_a^{\alpha_1\beta_1} G_b^{\alpha_2\mu} \tilde G_{c,\mu}^{\beta_2} - \{ \alpha_2 \leftrightarrow \beta_2 \} ] ,
\\
\\ \nonumber \tilde \xi_2^{\alpha_1\alpha_2,\beta_1\beta_2} &=& d^{abc} \mathcal{S}[ g_s^3 G_a^{\alpha_1\beta_1} \tilde G_b^{\alpha_2\mu} G_{c,\mu}^{\beta_2} - \{ \alpha_2 \leftrightarrow \beta_2 \} ] ,
\\
\\ \xi_3^{\cdots} &=& d^{abc} \mathcal{S}[ g_s^3 G_a^{\alpha_1\beta_1} G_{b}^{\alpha_2\beta_2} G_c^{\alpha_3\beta_3} ] \, ,
\\ \tilde \xi_3^{\cdots} &=& d^{abc} \mathcal{S}[ g_s^3 \tilde G_a^{\alpha_1\beta_1} \tilde G_{b}^{\alpha_2\beta_2} \tilde G_c^{\alpha_3\beta_3}] \, ,
\end{eqnarray}
where $d^{abc}$ is the totally symmetric $SU(3)_C$ structure constant. Their sum rule equations are:
\begin{eqnarray}
&& \Pi_{|{\rm GGG};1^{+-}\rangle}(s_0, M_B^2)
\label{eq:GGGope1pm}
\\ \nonumber &=& \int_0^{s_0} \Big( \frac{4 \alpha_s^3}{81 \pi} s^4 + \frac{10\alpha_s^2 \langle g_s^2 GG \rangle}{9} s^2
\\ \nonumber && ~~~~~ + \frac{35\alpha_s^3 \langle g_s^2 GG \rangle}{36\pi} s^2 + \frac{5\alpha_s^2 \langle g_s^3 G^3 \rangle}{27} s \Big) e^{-s/M_B^2} ds  ,
\\[2mm]
&& \Pi_{|{\rm GGG};2^{+-}\rangle}(s_0, M_B^2)
\label{eq:GGGope2pm}
\\ \nonumber &=& \int_0^{s_0} \Big( \frac{\alpha_s^3}{324 \pi} s^4 - \frac{5\alpha_s^2 \langle g_s^2 GG \rangle}{108} s^2
\\ \nonumber && ~~~~~ + \frac{15\alpha_s^3 \langle g_s^2 GG \rangle}{128\pi} s^2 + \frac{65\alpha_s^2 \langle g_s^3 G^3 \rangle}{216} s \Big) e^{-s/M_B^2} ds  ,
\\[2mm]
&& \Pi_{|{\rm GGG};3^{+-}\rangle}(s_0, M_B^2)
\label{eq:GGGope3pm}
\\ \nonumber &=& \int_0^{s_0} \Big( \frac{5\alpha_s^3}{2016 \pi} s^4 + \frac{\alpha_s^2 \langle g_s^2 GG \rangle}{16} s^2
\\ \nonumber && ~~~~~ - \frac{59\alpha_s^3 \langle g_s^2 GG \rangle}{512\pi} s^2 - \frac{\alpha_s^2 \langle g_s^3 G^3 \rangle}{2} s \Big) e^{-s/M_B^2} ds  ,
\\[2mm]
&& \Pi_{|{\rm GGG};1^{--}\rangle}(s_0, M_B^2)
\label{eq:GGGope1mm}
\\ \nonumber &=& \int_0^{s_0} \Big( \frac{4 \alpha_s^3}{81 \pi} s^4 - \frac{10\alpha_s^2 \langle g_s^2 GG \rangle}{9} s^2
\\ \nonumber && ~~~~~ + \frac{25\alpha_s^3 \langle g_s^2 GG \rangle}{36\pi} s^2 + \frac{35\alpha_s^2 \langle g_s^3 G^3 \rangle}{27} s \Big) e^{-s/M_B^2} ds  ,
\\[2mm]
&& \Pi_{|{\rm GGG};2^{--}\rangle}(s_0, M_B^2)
\label{eq:GGGope2mm}
\\ \nonumber &=& \int_0^{s_0} \Big( \frac{\alpha_s^3}{324 \pi} s^4 + \frac{5\alpha_s^2 \langle g_s^2 GG \rangle}{108} s^2
\\ \nonumber && ~~~~~ + \frac{15\alpha_s^3 \langle g_s^2 GG \rangle}{128\pi} s^2 + \frac{5\alpha_s^2 \langle g_s^3 G^3 \rangle}{24} s \Big) e^{-s/M_B^2} ds  ,
\\[2mm]
&& \Pi_{|{\rm GGG};3^{--}\rangle}(s_0, M_B^2)
\label{eq:GGGope3mm}
\\ \nonumber &=& \int_0^{s_0} \Big( \frac{5\alpha_s^3}{2016 \pi} s^4 - \frac{\alpha_s^2 \langle g_s^2 GG \rangle}{16} s^2
\\ \nonumber && ~~~~~ - \frac{49\alpha_s^3 \langle g_s^2 GG \rangle}{1536\pi} s^2 - \frac{11\alpha_s^2 \langle g_s^3 G^3 \rangle}{432} s \Big) e^{-s/M_B^2} ds  .
\end{eqnarray}
The above three-gluon glueball currents of $C=-$ have been systematically studied in Ref.~\cite{Chen:2021cjr}, but there we did not calculate the Feynman diagrams depicted in Figs.~\ref{fig:feynman}($c-3,c-4,c-5$). Similar to Eqs.~(\ref{eq:GGope0pp})-(\ref{eq:GGGope2mp}), we find all the $D=8$ terms proportional to $\langle g_s^2 GG \rangle^2$ vanish, so the convergence of the above OPE series are also quite good.

We shall use the above sum rule equations to perform numerical analyses in the next section.

%
\section{Numerical analyses}
\label{sec:numerical}
%

In this section we perform numerical analyses using the sum rules given in Eqs.~(\ref{eq:GGope0pp})-(\ref{eq:GGGope2mp}) and Eqs.~(\ref{eq:GGGope1pm})-(\ref{eq:GGGope3mm}). The glueball mass $M_X$ depends significantly on the gluon condensates $\langle g_s^2GG\rangle$ and $\langle g_s^3G^3\rangle$, both of which are still not precisely known. In the present study we use the following values for these parameters~\cite{Narison:2011xe,Narison:2018dcr}:
\begin{eqnarray}
\langle \alpha_s GG\rangle &=& (6.35 \pm 0.35) \times 10^{-2} \mbox{ GeV}^4 \, ,
\label{eq:condensate2}
\\ \nonumber \langle g_s^3G^3\rangle &=& \langle \alpha_s GG\rangle \times (8.2 \pm 1.0) \mbox{ GeV}^2 \, .
\end{eqnarray}
Besides, we use the following value for the strong coupling constant at the QCD scale $\Lambda_{\rm QCD} = 300$~MeV~\cite{pdg}:
%
\begin{equation}
\alpha_s(Q^2) = {4\pi \over 11 \ln(Q^2/\Lambda_{\rm QCD}^2)} \, .
\end{equation}
%

\begin{figure}[hbt]
\begin{center}
\includegraphics[width=0.45\textwidth]{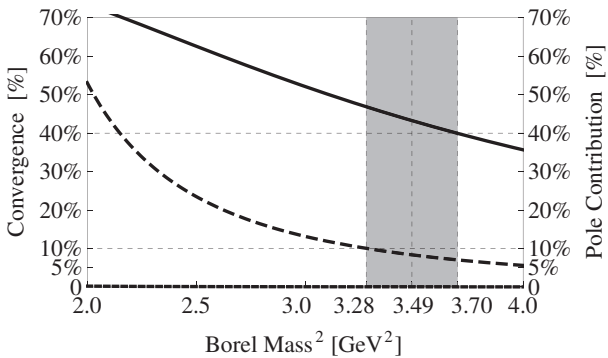}
\caption{CVG$_A$ (short-dashed curve, defined in Eq.~(\ref{eq:convergence1})), CVG$_B$ (long-dashed curve, defined in Eq.~(\ref{eq:convergence2})), and PC (solid curve, defined in Eq.~(\ref{eq:pole})) as functions of the Borel mass $M_B$. The current $\tilde J_0$ is used here when setting $s_0 = 9.0$~GeV$^2$.}
\label{fig:cvgpole}
\end{center}
\end{figure}

We still take the current $\tilde J_0$ as an example, and use Eq.~(\ref{eq:LSR}) to calculate the mass of $|{\rm GG};0^{-+}\rangle$. It depends on two free parameters, the Borel mass $M_B$ and the threshold value $s_0$. We use two criteria to determine the Borel window. The first criterion is to insure the convergence of OPE by requiring a) the $\alpha_s^2 \times g_s^2$ term $\alpha_s^2 \langle g_s^2 GG \rangle$ to be less than 5\%, and b) the $D=6$ term $\alpha_s \langle g_s^3 G^3 \rangle$ to be less than 10\%:
\begin{eqnarray}
\mbox{CVG}_A &\equiv& \left|\frac{ \Pi^{g_s^{n=6}}(s_0, M_B^2) }{ \Pi(s_0, M_B^2) }\right| \leq 5\% \, ,
\label{eq:convergence1}
\\
\mbox{CVG}_B &\equiv& \left|\frac{ \Pi^{{\rm D=6}}(s_0, M_B^2) }{ \Pi(s_0, M_B^2) }\right| \leq 10\% \, .
\label{eq:convergence2}
\end{eqnarray}
As shown in Fig.~\ref{fig:cvgpole} using the dashed curves, we determine the lower limit of $M_B$ to be $M_B^2 \geq 3.28$~GeV$^2$ when setting $s_0 = 9.0$~GeV$^2$.

The above condition is the cornerstone of a reliable sum rule analysis, where we have taken into account two terms because the OPE is expanded in two directions: the dimension of condensates and the coupling constant $g_s$. Eqs.~(\ref{eq:convergence1}) and (\ref{eq:convergence2}) are for two-gluon glueball currents, and the conditions for three-gluon glueball currents are
\begin{eqnarray}
\mbox{CVG}^\prime_A &\equiv& \left|\frac{ \Pi^{g_s^{n=8}}(s_0, M_B^2) }{ \Pi(s_0, M_B^2) }\right| \leq 5\% \, ,
\label{eq:convergence3}
\\
\mbox{CVG}^\prime_B &\equiv& \left|\frac{ \Pi^{{\rm D=6}}(s_0, M_B^2) }{ \Pi(s_0, M_B^2) }\right| \leq 10\% \, .
\label{eq:convergence4}
\end{eqnarray}

The second criterion is to insure the one-pole-dominance assumption by requiring the pole contribution (PC) to be larger than 40\%:
\begin{equation}
\mbox{PC} \equiv \left|\frac{ \Pi(s_0, M_B^2) }{ \Pi(\infty, M_B^2) }\right| \geq 40\% \, .
\label{eq:pole}
\end{equation}
As shown in Fig.~\ref{fig:cvgpole} using the solid curve, we determine the upper limit of $M_B$ to be $M_B^2 \leq 3.70$~GeV$^2$ when setting $s_0 = 9.0$~GeV$^2$.

\begin{figure*}[hbt]
\begin{center}
\includegraphics[width=0.4\textwidth]{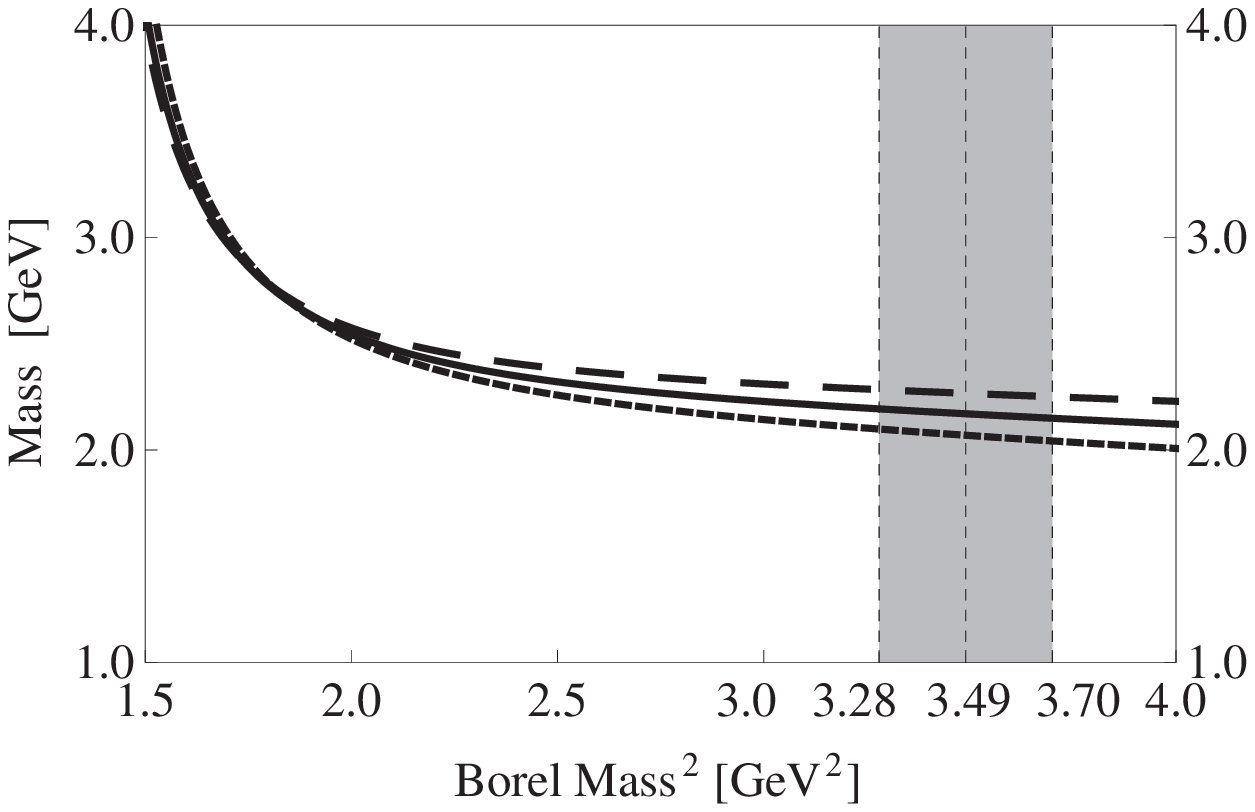}
~~~~~~~~~~~~
\includegraphics[width=0.4\textwidth]{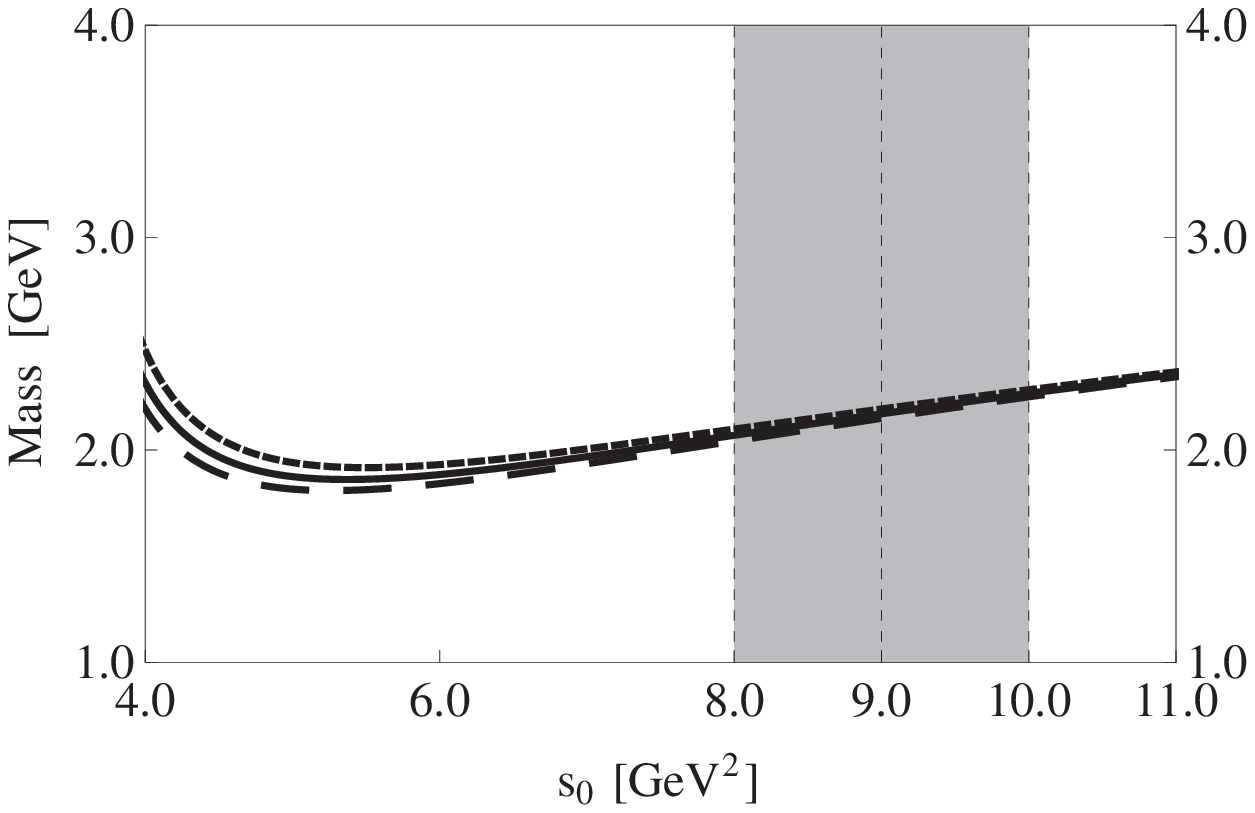}
\caption{
Mass of the two-gluon glueball $|{\rm GG};0^{-+}\rangle$ as a function of the Borel mass $M_B$ (left) and the threshold value $s_0$ (right), calculated using the current $\tilde J_0$.
In the left panel the short-dashed/solid/long-dashed curves are obtained by setting $s_0 = 8.0/9.0/10.0$ GeV$^2$, respectively.
In the right panel the short-dashed/solid/long-dashed curves are obtained by setting $M_B^2 = 3.28/3.49/3.70$ GeV$^2$, respectively.}
\label{fig:mass}
\end{center}
\end{figure*}

Altogether we determine the Borel window to be $3.28$~GeV$^2 \leq M_B^2 \leq 3.70$~GeV$^2$ for the fixed threshold value $s_0 = 9.0$~GeV$^2$. Then we redo the same procedures by changing $s_0$, and find that there exist non-vanishing Borel windows as long as $s_0 \geq s^{\rm min}_0 = 8.2$~GeV$^2$. Accordingly, we choose $s_0$ to be slightly larger, and determine our working regions to be $8.0$~GeV$^2 \leq s_0 \leq 10.0$~GeV$^2$ and $3.28$~GeV$^2 \leq M_B^2 \leq 3.70$~GeV$^2$, where we calculate the mass of $|{\rm GG};0^{-+}\rangle$ to be
\begin{equation}
M_{|{\rm GG};0^{-+}\rangle} = 2.17\pm0.11{\rm~GeV} \, .
\end{equation}
Its central value corresponds to $M_B^2=3.49$ GeV$^2$ and $s_0 = 9.0$ GeV$^2$, and its uncertainty comes from the threshold value $s_0$, the Borel mass $M_B$, and the gluon condensates listed in Eqs.~(\ref{eq:condensate2}).

We show $M_{|{\rm GG};0^{-+}\rangle}$ in the left panel of Fig.~\ref{fig:mass} as a function of the Borel mass $M_B$, and find it quite stable inside the Borel window $3.28$~GeV$^2 \leq M_B^2 \leq 3.70$~GeV$^2$. We also show it in the right panel of Fig.~\ref{fig:mass} as a function of the threshold value $s_0$. We find there exists a mass minimum around $s_0 \sim 5$~GeV$^2$, and the $s_0$ dependence is weak and acceptable inside the working region $8.0$~GeV$^2 \leq s_0 \leq 10.0$~GeV$^2$.

\begin{table*}[hbt]
\begin{center}
\renewcommand{\arraystretch}{1.5}
\caption{QCD sum rule results of two- and three-gluon glueballs.}
\begin{tabular}{c | c | c | c | c | c | c}
\hline\hline
\multirow{2}{*}{~~Glueball~~} & \multirow{2}{*}{~~~~~~Current~~~~~~} & \multirow{2}{*}{~$s_0^{min}~[{\rm GeV}^2]$~} & \multicolumn{2}{c|}{Working Regions} & \multirow{2}{*}{~~Pole~[\%]~~} & \multirow{2}{*}{~~Mass~[GeV]~~}
\\ \cline{4-5} & & & ~~$s_0~[{\rm GeV}^2]$~~ & ~~$M_B^2~[{\rm GeV}^2]$~~ &
\\ \hline \hline
$|{\rm GG};0^{++}\rangle$   & $J_0$                                            & 7.8  & $9.0\pm1.0$  & $3.70$--$4.19$   & $40$--$48$ & $1.78^{+0.14}_{-0.17}$
\\
$|{\rm GG};2^{++}\rangle$   & $J_2^{\alpha_1\alpha_2,\beta_1\beta_2}$          & 8.5  & $10.0\pm1.0$ & $3.99$--$4.60$   & $40$--$50$ & $1.86^{+0.14}_{-0.17}$
\\
$|{\rm GG};0^{-+}\rangle$   & $\tilde J_0$                                     & 8.2  & $9.0\pm1.0$  & $3.28$--$3.70$   & $40$--$47$ & $2.17^{+0.11}_{-0.11}$
\\
$|{\rm GG};2^{-+}\rangle$   & $\tilde J_2^{\alpha_1\alpha_2,\beta_1\beta_2}$   & 8.1  & $10.0\pm1.0$ & $3.27$--$4.20$   & $40$--$55$ & $2.24^{+0.11}_{-0.11}$
\\ \hline
$|{\rm GGG};0^{++}\rangle$  & $\eta_0$                                                & 31.6 & $33.0\pm3.0$ & $7.25$--$7.61$   & $40$--$44$ & $4.46^{+0.17}_{-0.19}$
\\
$|{\rm GGG};2^{++}\rangle$  & $\eta_2^{\alpha_1\alpha_2,\beta_1\beta_2}$              & 16.0 & $35.0\pm3.0$ & $4.77$--$9.04$   & $40$--$90$ & $4.18^{+0.19}_{-0.42}$
\\
$|{\rm GGG};0^{-+}\rangle$  & $\tilde \eta_0$                                         & 17.0 & $33.0\pm3.0$ & $4.48$--$8.13$   & $40$--$88$ & $4.13^{+0.18}_{-0.36}$
\\
$|{\rm GGG};2^{-+}\rangle$  & $\tilde \eta_2^{\alpha_1\alpha_2,\beta_1\beta_2}$       & 33.1 & $35.0\pm3.0$ & $8.10$--$8.53$   & $40$--$44$ & $4.29^{+0.20}_{-0.22}$
\\ \hline
$|{\rm GGG};1^{+-}\rangle$  & $\xi_1^{\alpha\beta}$                                            &  9.0 & $34.0\pm4.0$ & $3.16$--$9.09$ & $40$--$99$ & $4.01^{+0.26}_{-0.95}$
\\
$|{\rm GGG};2^{+-}\rangle$  & $\xi_2^{\alpha_1\alpha_2,\beta_1\beta_2}$                        & 32.7 & $35.0\pm4.0$ & $7.53$--$8.09$ & $40$--$46$ & $4.42^{+0.24}_{-0.29}$
\\
$|{\rm GGG};3^{+-}\rangle$  & $\xi_3^{\alpha_1\alpha_2\alpha_3,\beta_1\beta_2\beta_3}$         & 30.2 & $33.0\pm4.0$ & $7.69$--$8.40$ & $40$--$47$ & $4.30^{+0.23}_{-0.26}$
\\
$|{\rm GGG};1^{--}\rangle$  & $\tilde \xi_1^{\alpha\beta}$                                     & 31.2 & $34.0\pm4.0$ & $5.81$--$6.77$ & $40$--$51$ & $4.91^{+0.20}_{-0.18}$
\\
$|{\rm GGG};2^{--}\rangle$  & $\tilde \xi_2^{\alpha_1\alpha_2,\beta_1\beta_2}$                 & 19.7 & $36.0\pm4.0$ & $5.80$--$9.47$ & $40$--$81$ & $4.25^{+0.22}_{-0.33}$
\\
$|{\rm GGG};3^{--}\rangle$  & $\tilde \xi_3^{\alpha_1\alpha_2\alpha_3,\beta_1\beta_2\beta_3}$  & 35.8 & $38.0\pm4.0$ & $6.15$--$7.22$ & $40$--$49$ & $5.59^{+0.33}_{-0.22}$
\\ \hline\hline
\end{tabular}
\label{tab:mass}
\end{center}
\end{table*}

Similarly, we use the sum rules given in Eqs.~(\ref{eq:GGope0pp})-(\ref{eq:GGGope2mp}) and Eqs.~(\ref{eq:GGGope1pm})-(\ref{eq:GGGope3mm}) to perform numerical analyses, and calculate masses of two- and three-gluon glueballs systematically. The obtained results are summarized in Table~\ref{tab:mass}, where we choose threshold values $s_0$ for two-gluon glueballs to be around $9\sim10$~GeV$^2$, and those for three-gluon glueballs to be around $33\sim38$~GeV$^2$.

%
\section{Summary and Discussions}
\label{sec:summary}
%

In this paper we study two- and three-gluon glueballs of $C=+$ using the method of QCD sum rules, including
\begin{itemize}

\item the two-gluon glueballs with the quantum numbers $J^{PC} = 0^{\pm+}$, $1^{-+}$, and $2^{\pm+}$;

\item the three-gluon glueballs with the quantum numbers $J^{PC} = 0^{\pm+}$, $1^{\pm+}$, and $2^{\pm+}$.

\end{itemize}
We systematically construct their interpolating currents, and find that all the spin-1 currents of $C=+$ vanish, suggesting that the ``ground-state'' spin-1 glueballs of $C=+$ do not exist within the relativistic framework. This behavior is consistent with Lattice QCD calculations~\cite{Chen:2005mg,Mathieu:2008me,Meyer:2004gx,Gregory:2012hu}.

We use spin-0 and spin-2 glueball currents to perform QCD sum rule analyses, and calculate masses of their corresponding spin-0 and spin-2 glueballs. All these spin-2 currents have four Lorentz indices with certain symmetries, so that they couple to both positive- and negative-parity glueballs, which need to be further separated at the hadron level. We refer to Ref.~\cite{Chen:2021cjr} for detailed discussions.

\begin{table*}[hbt]
\begin{center}
\renewcommand{\arraystretch}{1.6}
\caption{Masses of two- and three-gluon glueballs, in units of GeV. Our QCD sum rule results are listed in the 2nd column. Lattice QCD results are listed in the 3rd-6th columns, taken from Refs.~\cite{Chen:2005mg,Mathieu:2008me,Meyer:2004gx} (quenched) and Ref.~\cite{Gregory:2012hu} (unquenched).}
\begin{tabular}{c | c | c | c | c | c}
\hline\hline
~~~~Glueball~~~~  & ~~QCD sum rules~~         & ~~~~~~~~Ref.~\cite{Chen:2005mg}~~~~~~~~ & ~~~~~~~~Ref.~\cite{Mathieu:2008me}~~~~~~~~ & ~~~~~~~~Ref.~\cite{Meyer:2004gx}~~~~~~~~  & ~~~Ref.~\cite{Gregory:2012hu}~~~
\\ \hline \hline
$|{\rm GG};0^{++}\rangle$  & $1.78^{+0.14}_{-0.17}$ & $1.71 \pm 0.05 \pm 0.08$                & $1.73 \pm 0.05 \pm 0.08$                   & $1.48 \pm 0.03 \pm 0.07$                  & $1.80 \pm 0.06$
\\
$|{\rm GG};2^{++}\rangle$  & $1.86^{+0.14}_{-0.17}$ & $2.39 \pm 0.03 \pm 0.12$                & $2.40 \pm 0.03 \pm 0.12$                   & $2.15 \pm 0.03 \pm 0.10$                  & $2.62 \pm 0.05$
\\
$|{\rm GG};0^{-+}\rangle$  & $2.17^{+0.11}_{-0.11}$ & $2.56 \pm 0.04 \pm 0.12$                & $2.59 \pm 0.04 \pm 0.13$                   & $2.25 \pm 0.06 \pm 0.10$                  & --
\\
$|{\rm GG};2^{-+}\rangle$  & $2.24^{+0.11}_{-0.11}$ & $3.04 \pm 0.04 \pm 0.15$                & $3.10 \pm 0.03 \pm 0.15$                   & $2.78 \pm 0.05 \pm 0.13$                  & $3.46 \pm 0.32$
\\ \hline
$|{\rm GGG};0^{++}\rangle$ & $4.46^{+0.17}_{-0.19}$ & --                                      & $2.67 \pm 0.18 \pm 0.13$                   & $2.76 \pm 0.03 \pm 0.12$                  & $3.76 \pm 0.24$
\\
$|{\rm GGG};2^{++}\rangle$ & $4.18^{+0.19}_{-0.42}$ & --                                      & --                                         & $2.88 \pm 0.10 \pm 0.13$                  & --
\\
$|{\rm GGG};0^{-+}\rangle$ & $4.13^{+0.18}_{-0.36}$ & --                                      & $3.64 \pm 0.06 \pm 0.18$                   & $3.37 \pm 0.15 \pm 0.15$                  & $4.49 \pm 0.59$
\\
$|{\rm GGG};2^{-+}\rangle$ & $4.29^{+0.20}_{-0.22}$ & --                                      & --                                         & $3.48 \pm 0.14 \pm 0.16$                  & --
\\ \hline
$|{\rm GGG};1^{+-}\rangle$ & $4.01^{+0.26}_{-0.95}$ & $2.98 \pm 0.03 \pm 0.14$                & $2.94 \pm 0.03 \pm 0.14$                   & $2.67 \pm 0.07 \pm 0.12$                  & $3.27 \pm 0.34$
\\
$|{\rm GGG};2^{+-}\rangle$ & $4.42^{+0.24}_{-0.29}$ & $4.23 \pm 0.05 \pm 0.20$                & $4.14 \pm 0.05 \pm 0.20$                   & --                                        & --
\\
$|{\rm GGG};3^{+-}\rangle$ & $4.30^{+0.23}_{-0.26}$ & $3.60 \pm 0.04 \pm 0.17$                & $3.55 \pm 0.04 \pm 0.17$                   & $3.27 \pm 0.09 \pm 0.15$                  & $3.85 \pm 0.35$
\\
$|{\rm GGG};1^{--}\rangle$ & $4.91^{+0.20}_{-0.18}$ & $3.83 \pm 0.04 \pm 0.19$                & $3.85 \pm 0.05 \pm 0.19$                   & $3.24 \pm 0.33 \pm 0.15$                  & --
\\
$|{\rm GGG};2^{--}\rangle$ & $4.25^{+0.22}_{-0.33}$ & $4.01 \pm 0.05 \pm 0.20$                & $3.93 \pm 0.04 \pm 0.19$                   & $3.66 \pm 0.13 \pm 0.17$                  & $4.59 \pm 0.74$
\\
$|{\rm GGG};3^{--}\rangle$ & $5.59^{+0.33}_{-0.22}$ & $4.20 \pm 0.05 \pm 0.20$                & $4.13 \pm 0.09 \pm 0.20$                   & $4.33 \pm 0.26 \pm 0.20$                  & --
\\ \hline \hline
\end{tabular}
\label{tab:comparison}
\end{center}
\end{table*}

We summarize the obtained results in Table~\ref{tab:comparison}, which are compared with the Lattice QCD results obtained using non-relativistic glueball operators~\cite{Chen:2005mg,Mathieu:2008me,Meyer:2004gx,Gregory:2012hu}. For completeness, we also reanalysis the results of $C=-$ three-gluon glueballs (also called as odderons), which have been previously studied in Ref.~\cite{Chen:2021cjr}. Therefore, a rather complete QCD sum rule study have been done on the lowest-lying glueballs composed of two or three valence gluons. We find that our QCD sum rule results are generally consistent with unquenched Lattice QCD results~\cite{Gregory:2012hu}.

To end this paper, we briefly discuss possible decay patterns of two- and three-gluon glueballs. The two-gluon glueballs can decay after exciting two quark-antiquark pairs, and recombine into two mesons. However, it is rather difficult to differentiate them from standard $q \bar q$ states. The three-gluon glueballs can decay after exciting three quark-antiquark pairs, and recombine into three mesons. Their possible decay patterns are:
\begin{eqnarray}
\nonumber 0^{-+} &\to& ~~~~~~~~\,VVP,VVV~~~~~~~~~(S\mbox{-wave}) \, ,
\\
\nonumber 0^{++} &\to& ~~~~\,VPP,VVP,VVV~~~~~(P\mbox{-wave}) \, ,
\\
\nonumber 1^{--} &\to& ~~~~\,VPP,VVP,VVV~~~~~(S\mbox{-wave}) \, ,
\\
\nonumber 1^{+-} &\to& PPP,VPP,VVP,VVV~(P\mbox{-wave}) \, ,
\\
\nonumber 2^{-\pm} &\to& ~~~~~~~~\,VVP,VVV~~~~~~~~~(S\mbox{-wave}) \, ,
\\
\nonumber 2^{+\pm} &\to& ~~~~VPP,VVP,VVV~~~~~(P\mbox{-wave}) \, ,
\\
\nonumber 3^{--} &\to& ~~~~~~~~~~~~\,VVV~~~~~~~~~~~~~(S\mbox{-wave}) \, ,
\\
\nonumber 3^{+-} &\to& ~~~~~~~~VVP,VVV~~~~~~~~~(P\mbox{-wave}) \, ,
\end{eqnarray}
where $P$ and $V$ denote light vector and pseudoscalar mesons, respectively.
Considering their limited decay patterns, the $J^{PC} = 0^{-+}/2^{-\pm}/3^{\pm-}$ three-gluon glueballs may have relatively smaller widths, and we propose to search for them in their $VVV$ and $VVP$ decay channels in future BESIII, GlueX, LHC, and PANDA experiments.


%
\section*{Acknowledgments}
%

We thank Zi-Yang Lin very much for helping us prove that all the spin-1 two- and three-gluon glueball currents of $C=+$ vanish.
We thank the anonymous PRD referee very much for checking our formulae and correcting one critical mistake.
This project is supported by
the National Natural Science Foundation of China under Grant No. 11722540, No.~11975033, No. 12075019, and No.~12070131001,
the National Key R$\&$D Program of China under Contracts No. 2020YFA0406400,
the Jiangsu Provincial Double-Innovation Program under Grant No.~JSSCRC2021488,
and
the Fundamental Research Funds for the Central Universities.

\appendix

\section{Spin-1 currents of $C=+$}
\label{app:spin1}

In this appendix we prove that the three spin-1 currents $J_1^{\alpha\beta}$, $\eta_1^{\alpha\beta}$, and $\tilde \eta_1^{\alpha\beta}$ all vanish. Their definitions are given in Eqs.~(\ref{def:J1}), (\ref{def:eta1}), and (\ref{def:eta1t}), respectively. For simplicity, we shall not differentiate the superscript and subscript in the following calculations.

Firstly, we investigate the current $J_1^{\alpha\beta}$. Due to the Lorentz invariant, we simply assume $\alpha = 0$ and $\beta = 1\cdots3$; besides, we need the Lorentz indices $\mu = 0/i$, $\rho = 0/k$, and $\sigma = 0/l$, with $i/k/l=1\cdots3$. We obtain:
\begin{eqnarray}
2 J_1^{0\beta} &=& 2 G_a^{0\mu} \tilde G_a^{\beta\mu} - \{ 0 \leftrightarrow \beta \} \, ,
\\ \nonumber &=& G_a^{0\mu} G_a^{\rho\sigma} \epsilon^{\beta\mu\rho\sigma} - G_a^{\beta\mu} G_a^{\rho\sigma} \epsilon^{0\mu\rho\sigma}
\\ \nonumber &=& G_a^{0i} G_a^{k0} \epsilon^{\beta i k0} + G_a^{0i} G_a^{0l} \epsilon^{\beta i 0l} - G_a^{\beta i} G_a^{k l} \epsilon^{0ikl} \, .
\end{eqnarray}
After interchanging $i \leftrightarrow k$, the first term turns out to be zero:
\begin{equation}
G_a^{0i} G_a^{k0} \epsilon^{\beta i k0} = G_a^{0k} G_a^{i0} \epsilon^{\beta k i0} = G_a^{0i} G_a^{k0} \epsilon^{\beta k i0} \rightarrow 0 \, .
\end{equation}
So does the second term. The third term is non-zero when $\beta = k$ or $\beta = l$. However, for the case $\beta = k$, we can interchange $i \leftrightarrow l$ and obtain (not sum over $\beta$ here):
\begin{equation}
G_a^{\beta i} G_a^{\beta l} \epsilon^{0i \beta l} = G_a^{\beta l} G_a^{\beta i} \epsilon^{0l \beta i} = G_a^{\beta i} G_a^{\beta l} \epsilon^{0l \beta i} \rightarrow 0 \, .
\end{equation}
So does the case $\beta = l$. Therefore, the third term is also zero, and the current $J_1^{\alpha\beta}$ vanishes.

Secondly, we investigate the current $\eta_1^{\alpha\beta}$:
\begin{eqnarray}
2 \eta_1^{\alpha\beta} &=& 2 f_{abc} \tilde G_a^{\mu\nu} G_b^{\mu\nu} G_c^{\alpha\beta}
\\ \nonumber &=& f_{abc} \epsilon^{\mu\nu\rho\sigma} G_a^{\rho\sigma} G_b^{\mu\nu} G_c^{\alpha\beta}
\\ \nonumber &=& f_{abc} \epsilon^{\mu\nu\rho\sigma} G_a^{\mu\nu} G_b^{\rho\sigma} G_c^{\alpha\beta}
\\ \nonumber &=& - f_{abc} \epsilon^{\mu\nu\rho\sigma} G_b^{\mu\nu} G_a^{\rho\sigma} G_c^{\alpha\beta}
\\ \nonumber &\rightarrow& 0 \, .
\end{eqnarray}
In the above expressions, we have consequently interchanged $\mu\nu \leftrightarrow \rho\sigma$ and $a \leftrightarrow b$. Similarly, we can prove the current $\tilde \eta_1^{\alpha\beta}$ to be zero.

One can construct more spin-1 three-gluon glueball currents of $C=+$, such as:
\begin{eqnarray}
\eta_1^{\prime\alpha\beta} &=& f_{abc} G_a^{\alpha\mu} G_b^{\mu\nu} G_c^{\nu\beta} - \{ \alpha \leftrightarrow \beta \} \, ,
\\
\tilde \eta_1^{\prime\alpha\beta} &=& f_{abc} G_a^{\alpha\mu} G_b^{\mu\nu} \tilde G_c^{\nu\beta} - \{ \alpha \leftrightarrow \beta \} \, .
\end{eqnarray}
It is straightforward to prove the former current $\eta_1^{\prime\alpha\beta}$ to be zero:
\begin{eqnarray}
\eta_1^{\prime\alpha\beta} &=& f_{abc} G_a^{\alpha\mu} G_b^{\mu\nu} G_c^{\nu\beta} - \{ \alpha \leftrightarrow \beta \}
\\ \nonumber &=& f_{abc} G_a^{\alpha\nu} G_b^{\nu\mu} G_c^{\mu\beta} - \{ \alpha \leftrightarrow \beta \}
\\ \nonumber &=& - f_{abc} G_a^{\beta\nu} G_b^{\nu\mu} G_c^{\mu\alpha} + \{ \alpha \leftrightarrow \beta \}
\\ \nonumber &=& f_{abc} G_c^{\beta\nu} G_b^{\nu\mu} G_a^{\mu\alpha} - \{ \alpha \leftrightarrow \beta \}
\\ \nonumber &=& - f_{abc} G_c^{\nu\beta} G_b^{\mu\nu} G_a^{\alpha\mu} - \{ \alpha \leftrightarrow \beta \}
\\ \nonumber &\rightarrow& 0 \, .
\end{eqnarray}
It is a bit tricky but one can still prove the latter current $\tilde \eta_1^{\prime\alpha\beta}$ to be zero, after explicitly writing out all its Lorentz indices. We have done this using the software {\it Mathematica}.

\end{document}